\documentclass[aps,prd,reprint,
superscriptaddress, 
nofootinbib,preprintnumbers]{revtex4-2}  
\usepackage[T1]{fontenc} 
\usepackage{enumitem}
\usepackage{bbm}
\usepackage[utf8]{inputenc}
\usepackage{amsmath}
\usepackage{amssymb}
\usepackage{pifont}
\usepackage{bm}
\usepackage{enumitem}
\usepackage{makecell}
\usepackage{subfig} 
\usepackage{tabularx,booktabs}
\newcolumntype{C}{>{\centering\arraybackslash}X}
\usepackage{graphicx}
\usepackage{multirow}
\usepackage[normalem]{ulem}
\usepackage{comment}
\usepackage{booktabs}
\usepackage{slashed}
\usepackage[table, svgnames, dvipsnames]{xcolor}
\usepackage{hyphenat} 
\usepackage{multirow,bigdelim}  
\usepackage{lipsum}
\usepackage{makecell}

\usepackage{tikz}
\usepackage[compat=1.1.0]{tikz-feynman}
\usetikzlibrary{calc}

\usetikzlibrary{shapes.misc}
\tikzset{
	>=latex,
    photon/.style={decorate, decoration={snake}, draw=black, thick},
    fermionnoarrow/.style={draw=black, postaction={decorate}, thick},
    scalar/.style={draw=black, postaction={decorate}, decoration={markings,mark=at position .55 with {\arrow{>}}}, thick, dashed},
    scalarnoarrow/.style={draw=black, postaction={decorate},  thick, dashed},
    fermion/.style={draw=black, postaction={decorate},decoration={markings,mark=at position .55 with {\arrow{>}}}, thick},
    gluon/.style={decorate, draw=black, decoration={coil,amplitude=4pt, segment length=5pt}, thick},
    vertex/.style={draw,shape=circle,fill=black,minimum size=3pt,inner sep=0pt},
    fillvertex/.style={draw,shape=circle,fill=black,minimum size=5pt,inner sep=0pt},
    openvertex/.style={draw,shape=circle,minimum size=51pt,inner sep=0pt},
    blob/.style={draw=black,shape=circle,fill=black,minimum size=6pt,inner sep=0pt},
    redvertex/.style={draw=red,shape=circle,fill=red,minimum size=3pt,inner sep=0pt},
    cross/.style={cross out, draw=black,thick, minimum size=5pt, inner sep=0pt, outer sep=0pt}
}

\usepackage[colorlinks=true
  ,urlcolor=emerald
  ,anchorcolor=emerald
  ,citecolor=emerald
  ,filecolor=emerald
  ,linkcolor=emerald
  ,menucolor=emerald
  ,linktocpage=true
  ,pdfproducer=medialab
  ,pdfa=true
]{hyperref}
\usepackage{cleveref}

\usepackage{tikz-feynman}
\tikzfeynmanset{compat=1.1.0}
\setlist[enumerate]{wide=0pt, widest=99,leftmargin=\parindent, labelsep=*}

\newdimen\arrowsize
\newdimen\mylw
\pgfkeys{/my arrows/chemeq/.style={draw,thick,double distance=2pt,onearc-onearc}}
\pgfkeys{/my arrows/size/.code={\pgfsetarrowoptions{onearc}{#1}}}
\def\myalw{.4pt}
\pgfarrowsdeclare{onearc}{onearc}{%
  \mylw=\myalw
  \pgfarrowsleftextend{-\pgfgetarrowoptions{onearc}-.5\mylw}
  \pgfarrowsrightextend{0pt}
}{%
  \pgfsetdash{}{0pt}
  \mylw=\pgflinewidth
  \pgfsetlinewidth{\myalw}
  \advance\arrowsize by.5\pgflinewidth
  \pgfpathmoveto{\pgfpoint{-\pgfgetarrowoptions{onearc}}{-\pgfgetarrowoptions{onearc}-.5\mylw}}%
  \pgfpatharc{180}{90}{\pgfgetarrowoptions{onearc}}
  \pgfusepathqstroke
}
\definecolor{brickred}{cmyk}{0, 0.9, 0.98, 0.4}
\definecolor{emerald}{cmyk}{1, 0.5, 0, 0}
\definecolor{gainsboroCMYK}{cmyk}{0, 0, 0, 0.14}

\makeatletter
\renewcommand\subsubsection{\@startsection{subsubsection}{3}{0pt}%
  {2.5ex plus 1ex minus .2ex}
  {1.5ex plus .5ex minus .2ex}
  {\centering\normalfont\normalsize\bf}} 
\makeatother

\begin{document}
\preprint{}

\title{\color{brickred}Generalised Casas-Ibarra Parametrisation for Majorana Neutrino Masses}

\author{Juan Herrero-García}
\email[E-mail: ]{juan.herrero@ific.uv.es}
\affiliation{Instituto de Física Corpuscular (CSIC-Universitat de València),
Parc Científic UV, C/Catedrático José Beltrán, 2, E-46980 Paterna, Spain}
\affiliation{Departament de Física Teòrica, Universitat de València, 46100 Burjassot, Spain}

\author{Simone Marciano}
\email[E-mail: ]{simone.marciano@ific.uv.es}
\affiliation{Instituto de Física Corpuscular (CSIC-Universitat de València),
Parc Científic UV, C/Catedrático José Beltrán, 2, E-46980 Paterna, Spain}
\affiliation{Departament de Física Teòrica, Universitat de València, 46100 Burjassot, Spain}

\author{Juan Racker}
\email[E-mail: ]{jracker@unc.edu.ar}
\affiliation{Universidad Nacional de C\'ordoba (UNC). Observatorio Astron\'omico de C\'ordoba (OAC). C\'ordoba, Argentina}
\affiliation{Consejo Nacional de Investigaciones Cient\'{\i}ficas y T\'ecnicas (CONICET), Instituto de Astronom\'{\i}a Te\'orica y Experimental (IATE), Laprida 854, X5000BGR, Córdoba, Argentina}

\author{Drona Vatsyayan}
\email[E-mail: ]{drona@physics.carleton.ca}
\affiliation{Instituto de Física Corpuscular (CSIC-Universitat de València),
Parc Científic UV, C/Catedrático José Beltrán, 2, E-46980 Paterna, Spain}
\affiliation{Department of Physics, Carleton University, Ottawa, ON K1S 5B6, Canada}


\begin{abstract} 
We present a simple and broadly applicable extension of the Casas–Ibarra parametrisation that captures the structure of all Majorana neutrino mass models. Building directly on the original formulation, our approach naturally accommodates additional degrees of freedom and provides a unified, minimal framework for parametrising the Yukawa sector. It significantly simplifies both analytical treatments and numerical scans, and can be universally applied to any Majorana neutrino mass model, regardless of the underlying dynamics. The approach also offers a unified framework for classifying neutrino mass models according to the structure of the neutrino mass matrix, which naturally motivates the proposal of an extended version of the Scotogenic Model. This classification scheme yields tree-level (loop-level) representative models: the seesaw (Scotogenic Model), the linear seesaw (the Generalised Scotogenic Model), and the linear plus inverse seesaw (the Extended Scotogenic Model). We provide ready-to-use explicit expressions for several well-known scenarios, including the Zee model where one of the Yukawa matrices is antisymmetric.
\end{abstract}

\maketitle



\section{Introduction}

The origin of the light neutrino masses remains one of the most interesting questions in particle physics. Several possible mechanisms have been proposed, one of the most compelling explanations being that neutrinos are Majorana particles, as hinted by the Weinberg operator \cite{Weinberg:1979sa}. One of the simplest high-energy completions are heavy right-handed neutrinos, such that light neutrino masses are generated by the (Type-I) seesaw mechanism~\cite{Minkowski:1977sc,Yanagida:1979as,Yanagida:1979gs,Glashow:1979nm,Gell-Mann:1979vob}. Other possibilities, like radiative neutrino mass models, have also been extensively explored~\cite{Ma:1998dn,Cai:2017jrq,Gargalionis:2020xvt}.

To study the phenomenology and do parameter scans of neutrino mass models, it is very useful to parametrise the Yukawa couplings that appear in the neutrino mass matrix in terms of light neutrino parameters\footnote{It is worth noting the existence of models in which lepton mixing is connected to 
quark mixing~\cite{Minakata_2004,Giarnetti:2024vgs,Harada_2006,PICARIELLO_2007,Chauhan_2007,Barranco:2010we,Zhang:2012pv,Zhukovsky:2019eoy}, without the 
need for Grand Unification. In such frameworks, the PMNS matrix is expressed in terms of the quark mixing parameters, so that the Yukawa couplings depend on both neutrino and quark observables, combined in specific ways depending on the model.
} (masses, mixings and phases) and other free parameters of the models. For this goal, several parametrisations have been proposed in the literature. In particular, this approach has been very successfully applied to the Type-I seesaw model using the Casas-Ibarra (CI) parametrisation~\cite{Casas:2001sr}, studied for special Yukawa textures in Ref.~\cite{Ibarra:2003xp}, simplified in the case of just two right-handed neutrinos in Ref.~\cite{Ibarra:2003up}, and generalised to include radiative corrections~\cite{Pilaftsis:1991ug} in~\cite{Lopez-Pavon:2015cga} (see also Refs.~\cite{Broncano:2003fq,Guo:2006qa,Xing:2007zj,Xing:2011ur,Heeck:2012fw,Xing:2024cyc} for other studies). In Refs.~\cite{Blennow:2011vn,Donini:2012tt}, other exact parametrisations for models with right-handed neutrinos, valid outside the seesaw limit, have been derived. For Type-II seesaw~\cite{Magg:1980ut,Mohapatra:1980yp,Lazarides:1980nt,Schechter:1980gr}, obtaining the (symmetric) Yukawa matrix in terms of neutrino masses is trivial, while for the Type-I+II seesaw a parametrisation has been proposed in Ref.~\cite{Akhmedov:2008tb}. Furthermore, parametrisations similar to the CI one for Type-III and inverse seesaw can be found in Refs.~\cite{Das:2012ze, Das:2020uer}, a modified CI parametrisation for the Krauss{-}Nasri{-}Trodden (KNT) model~\cite{Krauss:2002px} was given in Ref.~\cite{Cepedello:2020lul}, and a generalised CI parametrisation for an extension of the scotogenic model~\cite{Tao96, Ma:2006km} with several inert doublets was proposed in~\cite{Ahriche:2022bpx}.

It is worth noting that, although in many works the seesaw parameter space is explored by adopting the standard CI parametrisation -- where several degrees of freedom of the model are embedded in an orthogonal complex matrix -- this is not the only possible approach. 
Alternative formulations have been proposed in the literature, 
aiming to relate more directly the parameters of the high-energy Lagrangian to 
low-energy observables. 
In particular, Refs.~\cite{Davidson:2001zk,  Ellis:2002fe, Ibarra:2005qi} have shown that the parametrisation can be reformulated in terms of the hermitian combination $Y^\dagger Y$ of the Yukawa couplings, which provides a complementary perspective and can be particularly advantageous in the supersymmetric seesaw. Driven by the same motivation, another interesting parametrisation for the inverse seesaw was presented in~\cite{Kriewald:2024rlg}.

In addition, other formulations for the Yukawa structure in less minimal models have been proposed. In Ref.~\cite{Babu:2002uu}, a parametrisation of the $n_L \times n_L$ antisymmetric Yukawa matrix of the singly-charged scalar present in the Zee-Babu model~\cite{Babu:1988ki}, which appears quadratically, has been obtained using the fact that the latter has a zero eigenvalue and it has been extensively used in phenomenological studies \cite{AristizabalSierra:2006gb,Nebot:2007bc,Ohlsson:2009vk,Herrero-Garcia:2014hfa,Schmidt:2014zoa}. Further, Ref.~\cite{Felkl:2021qdn} also using this fact, obtained some restrictions on the antisymmetric Yukawa matrix of models in which it appears linearly, like in the Zee model~\cite{Zee:1980ai,Cheng:1980qt,Wolfenstein:1980sy} (see also Ref.~\cite{Chen:2025thp}). Finally, in Refs.~\cite{Cordero-Carrion:2018xre,Cordero-Carrion:2019qtu}, another general but more involved parametrisation for Majorana neutrino mass models has been proposed.  

In this work, we demonstrate that the CI parametrisation can be generalised to encompass all Majorana neutrino mass models, providing a unified framework for Majorana neutrino masses.  Although related ideas have appeared in the literature, a self-contained and comprehensive formulation has so far been lacking. To illustrate the straightforward applicability of this approach, we apply it to several prototypical examples, like the linear/inverse seesaw models and the Generalised Scotogenic Model. We also include the case in which one Yukawa matrix entering neutrino masses is antisymmetric, like in the Zee model, where we provide a completely explicit parametrisation in terms of some chosen free parameters. As a spin-off, our approach naturally leads us to propose a new version of the Scotogenic Model, which we term Extended Scotogenic Model. 

The rest of the paper is organised as follows. In Sec.~\ref{sec:general} we describe the general neutrino mass matrix and the generalised CI parametrisation.  In Sec.~\ref{sec:models} we \emph{follow} the form of the neutrino mass matrix and study typical model realisations of it, both at tree level and at one loop: \emph{i)} the seesaw and the Scotogenic Model; \emph{ii)} the linear seesaw and the Generalised Scotogenic Model; and \emph{iii)} the linear+inverse seesaw and the Extended Scotogenic Model. In Sec.~\ref{sec:Zee} we discuss models with extra symmetries and the application of the generalised CI parametrisation to the Zee Model. In Sec.~\ref{sec:CIgen} we discuss an approach to parametrise the Yukawas using the complete neutrino mass matrix in models with right-handed neutrinos. In App.~\ref{app:oldCI} we derive the standard CI in the seesaw limit using this approach.
Our conclusions are summarised in Sec.~\ref{sec:conc}. \\

\section{General neutrino mass matrix and parametrisation}\label{sec:general}
We are interested in the case where the $n_L=3$ light neutrinos $\nu_L$ acquire Majorana masses, 
\begin{equation}\label{eq:Maj}
       -\mathcal{L}_{\nu}= \frac{1}{2}\,\overline{\nu_L^c}\, m_\nu\, \nu_L+\text{ H.c.}\,.
\end{equation}
In general, the $n_L\times n_L$ light complex symmetric Majorana neutrino mass matrix may receive several contributions, either at tree level and/or from radiative corrections. If there are several contributions to light neutrino masses, the latter may be expressed in all generality as 
\begin{equation}\label{completemassmatrixa}
    m_\nu=\sum^N_{i,j=1}Y_i^T M_{ij}Y_j\,,
\end{equation}
where the complex Yukawa matrix $Y_i$ has dimensions $n_i\times n_L$ ($i=1, \dots, N$) and the complex mass matrix $M_{ij}=M^T_{ji}$ has dimensions $n_i\times n_j$. For simplicity, let us consider in the following the case of $N=2$, which in neutrino mass models corresponds to just two different sets of Yukawa couplings, $Y_1$ and $Y_2$. Then, the light neutrino mass matrix becomes:
\begin{align}\label{completemassmatrix2}
    m_\nu&=Y_1^T M_{11}Y_1+Y_1^T M_{12}Y_2\nonumber\\
    &+Y_2^T M_{21}Y_1+Y_2^T M_{22}Y_2\,.
\end{align}
The latter form may be mapped onto neutrino mass–model parameters, which we exploit in Sec.~\ref{sec:models}. The sum of these contributions can be rewritten as
\begin{equation}\label{ECIMassMatrixa}
    m_\nu=\mathcal{Y}^T\, \mathcal{M} \,\mathcal{Y}\,,
\end{equation}
where
\begin{equation} \label{eq:MMa}
    \mathcal{Y}=\begin{pmatrix}
        Y_1\\
        Y_2
    \end{pmatrix}\,,\quad\text{and}\quad 
    \mathcal{M}=\begin{pmatrix}
        M_{11}&M_{12}\\
        M_{12}^T&M_{22}
    \end{pmatrix}\,,
\end{equation}
with $\mathcal{M}=\mathcal{M}^T$ being complex symmetric. Therefore, the neutrino mass matrix $m_\nu$ can always be expressed in the form of Eq.~\eqref{ECIMassMatrixa}, regardless of the underlying physics or the order at which the different mass contributions arise. Note that this clearly also holds for $N>2$, using appropriate block matrices. This has already been pointed out in App.~F of Ref.~\cite{Cordero-Carrion:2019qtu}.\footnote{Eq.~1 of Refs.~\cite{Cordero-Carrion:2018xre,Cordero-Carrion:2019qtu} presents the general Majorana neutrino mass matrix in a form that, while algebraically equivalent to ours, is less transparent when several diagonal and off-diagonal terms are present. In such cases - and when $y_1$ and $y_2$ in that equation are subject to specific constraints - a direct use of their master parametrisation does not follow automatically.}

To study the phenomenology, it is useful to express the Yukawa couplings in terms of light neutrino parameters and other free parameters. For Type-I seesaw, this may be done via the CI parametrisation~\cite{Casas:2001sr}, which can be used in the seesaw limit (see Sec.~\ref{sec:SS} below). However, in the generic case, before being able to apply the CI procedure~\cite{Casas:2001sr}, the extended mass matrix $\mathcal{M}$ has to be diagonalised via the Autonne-Takagi factorisation
\begin{equation} \label{eq:Vdiag}
    \mathcal{M} = V^T D_\mathcal{M} V \,,
\end{equation}
where $D_\mathcal{M}$ is a real diagonal matrix with non-negative entries and $V$ is a unitary matrix,  $V^\dagger V=\mathbb{I}_{n_1+n_2}$. Then exactly the same derivation as in Ref.~\cite{Casas:2001sr} can be applied to $V \mathcal{Y}$. Also by means of the Autonne-Takagi factorisation, the $n_L \times n_L$ light neutrino mass matrix may be written as
\begin{equation}\label{eq:Dm}
    D_m=\text{diag}(m_1,m_2,m_3)=U^T\, m_\nu\, U\,,
\end{equation}
with $U$ a $n_L \times n_L$ unitary matrix. Then we can write the \emph{generalised Casas-Ibarra} (GCI) parametrisation as
\begin{equation}  \label{eq:genCI}
\mathcal{Y}=V^\dagger D^{-1/2}_{\mathcal{M}}\,\mathcal{R}\,D_{\sqrt{m}}\,U^\dagger\,,
\end{equation}
where $ \mathcal{R}$ is an $(n_1+ n_2) \times n_L$ (semi)-orthogonal matrix, $\mathcal{R}^T\mathcal{R}=\mathbb{I}_{n_L}$, and we assume that there are no strictly vanishing eigenvalues in $D_{\mathcal{M}}$. Here the word \emph{generalised} highlights that, by writing the Majorana neutrino mass matrix as in Eq.~\eqref{ECIMassMatrixa} and diagonalising also $\mathcal{M}$, the structure of the CI parametrisation of Ref.~\cite{Casas:2001sr} actually holds for any Majorana neutrino mass model.\footnote{A generalised CI parametrisation was proposed in~\cite{Ahriche:2022bpx}, but in the context of one specific model, namely a scotogenic model with several inert doublets (rather than one as in~\cite{Tao96, Ma:2006km}), and using an orthogonal matrix in one of the diagonalisation steps, instead of the unitary matrix appearing in the Autonne-Takagi factorisation of Eq.~\eqref{eq:Vdiag}.}

Note that this expression is valid for any field basis. Indeed,
one can work in a basis where the charged lepton mass matrix is not diagonal, as discussed, e.g., in Refs.~\cite{Ibarra:2003xp,Ibarra:2003up}. In this case, the matrix $U$ can be written as 
$U=U_\text{PMNS}\; U_L$,
where $U_\text{PMNS}$ is the standard PMNS mixing matrix~\cite{Pontecorvo:1957cp,Pontecorvo:1957qd,Maki:1962mu,Pontecorvo:1967fh} and $U_L$ accounts for the charged lepton mixing.\footnote{The choice of mass ordering (normal or inverted) becomes relevant only at this point, via the parametrisation of the PMNS matrix, $U_\text{PMNS}$, and the order of the eigenvalues in $D_m$.} It is also possible to operate in a basis where, e.g., the Majorana mass matrix of the right-handed neutrinos in the Type-I seesaw is not diagonal, as in Ref.~\cite{Ibarra:2003up}. However, we emphasise that in the GCI parametrisation proposed in Eq.~\eqref{eq:genCI}, the matrix $V$ is the one that diagonalises the extended mass matrix $\mathcal{M}$ and it is not necessarily related to a change of basis, as in~\cite{Ibarra:2003up}.

We summarise the important result discussed in this section:
\begin{equation*}
\fbox{\parbox{0.9\linewidth}{
Any $n_L \times n_L$ Majorana neutrino mass matrix, which may have an arbitrary number of contributions,
\begin{equation}\label{completemassmatrix}
    m_\nu=\sum^N_{i,j=1}Y_i^T M_{ij}Y_j\,,
\end{equation}
with Yukawas $Y_i$ of dimensions $n_i \times n_L$, may be always written in the usual \emph{seesaw} way,
\begin{equation} \label{eq:mnuSS}
\boxed{m_\nu = {\mathcal Y}^T\,{\mathcal M}\,{\mathcal Y}}
\end{equation}
with $\mathcal{Y}$ an $(n_1+\ldots +n_N) \times n_L$ Yukawa matrix, and $\mathcal{M}$ an $(n_1+\ldots +n_N) \times (n_1+\ldots +n_N)$ complex symmetric matrix. Therefore, the seesaw expression is the most general form for any Majorana (neutrino) mass matrix. A \emph{generalised Casas-Ibarra} (GCI) parametrisation may be used,
\begin{equation} \label{eq:CIgen}
\mathcal{Y}=V^\dagger D^{-1/2}_{\mathcal{M}}\,\mathcal{R}\,D_{\sqrt{m}}\,U^\dagger\,~\text{GCI\,\,parametrisation}\,,
\end{equation}
where $D_\mathcal{M}=V^\ast \mathcal{M} V^\dagger$ and $\mathcal{R}$ is a complex orthogonal matrix, $\mathcal{R}^T\mathcal{R}=\mathbb{I}_{n_L}$.
This is a simple and well-suited parametrisation for numerical scans relevant to phenomenological analyses.}}
\end{equation*}

A word of caution is in order. Note that, if the Yukawa matrices are not general, extra conditions on the matrix $\mathcal{R}$ may need to be imposed. For instance, this is the case of antisymmetric Yukawa matrices, $f=-f^T$, which appear for instance in the Zee (linear in the antisymmetric Yukawa)~\cite{Zee:1985id} and Zee-Babu models (quadratic in the antisymmetric Yukawa coupling)~\cite{Babu:1988ki}, see also Ref.~\cite{Cheng:1980qt}. For the Zee-Babu case, another parametrisation which uses the fact that an antisymmetric matrix has a zero eigenvalue is much more straightforward to use for phenomenological studies~\cite{Babu:2002uu,AristizabalSierra:2006gb,Nebot:2007bc,Herrero-Garcia:2014hfa,Felkl:2021qdn}, while for the Zee model~\cite{Zee:1985id} (see Refs.~\cite{Herrero-Garcia:2017xdu,Heeck:2023iqc,Babu:2019mfe} for detailed studies of the phenomenology), in Sec.~\ref{sec:Zee} we derive a parametrisation of $\mathcal{R}$ that ensures the antisymmetric condition.

\subsection{Specific form of the (semi-)orthogonal Casas-Ibarra $\mathcal{R}$ matrix}
The key point of the CI parametrisation lies in embedding all the free parameters of the neutrino sector -- except for the eigenvalues of the $\mathcal{M}$ matrix -- into the $\mathcal{R}$ matrix, thereby making the phenomenological analysis simple and straightforward. The matrix $\mathcal{R}$ may, in general, take a non-square form depending on the dimensions of the Yukawa coupling matrices. To build a semi-orthogonal $(n_1+n_2) \times n_L$ $\mathcal{R}$ matrix we can start from the special orthogonal group $SO(n)$. It is well known that its dimension is $d=n(n-1)/2$, so that a matrix $\mathcal{R}\in SO(n)$ can be parametrised with $d$ rotation angles as the product of plane rotations in $d$ coordinates, as follows:
\begin{equation}
    \mathcal{R}=\prod_{j=2}^{n} \prod_{i=1}^{j-1}R_{ij}\,,
\end{equation}
with
\begin{equation}
    \left[R_{ij}\right]_{ab}=\begin{cases}
        c_{ij}\quad a=b=i, \text{ or } a=b=j\,,\\
        s_{ij}\quad a=i, b=j\,,\\
        -s_{ij}\quad a=j, b=i\,,\\
        1\quad a=b\not\in\{i,j\}\,,\\
        0\quad \text{otherwise}\,,
    \end{cases}
\end{equation}
where $c_{ij}=\cos\theta_{ij}$ and $s_{ij}=\sin\theta_{ij}$.
A procedure to obtain the semi-orthogonal matrix $\mathcal{R}$ with dimensions $(n_1 + n_2) \times n_L$ is to start from a square matrix belonging to $SO(n_L)$ if $n_1 + n_2 < n_L$, or to $SO(n_1 + n_2)$ otherwise, and then reduce its dimension so as to reach the desired $(n_1 + n_2) \times n_L$ form. 
In general, a $\mathcal{R}_{n\times m}$ semi-orthogonal matrix can be obtained as:
\begin{equation}\label{eq:Rnm_cases}
\mathcal{R}_{\,n\times m} =
\begin{cases}
\displaystyle
\left(
\begin{array}{c|c}
\mathbb{I}_{n} & 0_{\,n\times (m-n)}
\end{array}
\right) \, \mathcal{R}_{\,m\times m}, & n < m, \\[2ex]
\displaystyle
\mathcal{R}_{\,n\times n} \,
\begin{pmatrix}
\mathbb{I}_{m} \\ \hline
0_{(n-m)\times m}
\end{pmatrix}, & n > m,
\end{cases}
\end{equation}
where $0_{\,pq}$ is a $p\times q$ matrix of zeroes.
As an example, let us consider the case $n_1=n_2=1$ and substitute explicitly $n_L=3$. The most general $\mathcal{R}\in SO(3)$ is
\begin{equation}
\mathcal{R} =
\begin{pmatrix}
c_{12} c_{13} & s_{12} c_{13} & s_{13} \\
- s_{12} c_{23} - c_{12} s_{23} s_{13} &
c_{12} c_{23} - s_{12} s_{23} s_{13} &
s_{23} c_{13} \\
s_{12} s_{23} - c_{12} c_{23} s_{13} &
- c_{12} s_{23} - s_{12} c_{23} s_{13} &
c_{23} c_{13}
\end{pmatrix}\,,
\label{general3x3R}
\end{equation}
and we can obtain the desired $\mathcal{R}_{2\times 3}$ following Eq.~\eqref{eq:Rnm_cases}:
\begin{equation}\label{eq:cut3x3}
     \mathcal{R}_{2\times 3}\equiv\left(
\begin{array}{c|c}
\mathbb{I}_{2} & 0_{\,2\times 1}
\end{array}
\right)\times\mathcal{R}_{3\times 3}\,.
\end{equation}

\section{Implementations of the mass matrix and the Generalised Casas-Ibarra parametrisation} \label{sec:models}

In the following, we study in turn tree-level and loop-level implementations of diagonal, off-diagonal and general neutrino mass matrices. We consider specific models that are widely discussed in the literature and illustrate how to implement for them the GCI parametrisation. In all cases we start from a basis in which the $M_{ij}$ matrices defined above are diagonal, and we use the notation $\overline{M}_i$ for the eigenvalues of $\mathcal{M}$. Also note that, when working in the basis in which the charged lepton mass matrix is diagonal, the matrix $U$ appearing in all the expressions we give below coincides with the PMNS matrix, $U_\text{PMNS}$. For the latest values of the oscillation parameters, see Refs.~\cite{Esteban:2024eli,NuFIT60}.

\subsection{Diagonal mass matrix} \label{sec:diag}
In some cases, there may only be one source of neutrino masses. In the following we study the case in which the diagonal element $M_{11}$ is non-zero,  $M_{11}\neq 0$, while the rest of elements are zero, $M_{12}=M_{21}=M_{22}=0$.

\subsubsection{Tree level: the Seesaw case} \label{sec:SS}

The most famous example of $M_{11}\neq 0$ at tree level is the usual seesaw limit of models with extra right-handed neutrinos. If \(n_R=n_1\) right-handed neutrinos $N$ are added to the SM, the Yukawa and mass terms in the Lagrangian read \cite{Minkowski:1977sc,Yanagida:1979as,Gell-Mann:1979vob,Yanagida:1979gs,Glashow:1979nm}:
\begin{equation} \label{eq:SS}
    -\mathcal{L}_\text{SS}= \overline{L}Y \tilde H N+ \frac{1}{2}\,\overline{N^c}\, m_R\, N+\text{ H.c.}\,,
\end{equation}
where $m_{\rm R}$ is a $n_R\times n_R$ complex Majorana mass matrix which, without loss of generality, can be chosen to be diagonal with positive eigenvalues, and $Y$ is a $n_L \times  n_R$ complex Yukawa matrix. Here $L=(\nu_L, e_L)^T$ are the left-handed lepton doublets.\footnote{This flavour structure also applies, for instance, to Type-III seesaw and Model $A_1$ in Ref.~\cite{Giarnetti:2023osf}.} We denote by $H$ the SM Higgs doublet, which in the unitary gauge after electroweak symmetry breaking is given by
\begin{equation}
H \equiv (0, (h + v)/\sqrt{2})^T\,,
\end{equation}
with $ v = 246 $ GeV the vacuum expectation value (VEV) and $ h $ the Higgs boson. In the basis $(\nu_L, N^c)$, after Electroweak Spontaneous Symmetry Breaking (EW SSB), the complete \((n_L+n_R)\times(n_L+n_R)\) neutral fermion mass matrix reads
\begin{equation}\label{eq:MMSS2}
M_\nu=\begin{pmatrix}
0_{n_L\times n_L} & m_D\\[4pt]
m_D^T & m_R
\end{pmatrix},
\end{equation}
where \(m_D=Y\,v/\sqrt2\) (so \(m_D\) is \(n_L\times n_R\)). In the seesaw limit, $m_D \ll m_R$, Eq.~\eqref{eq:MMSS2} may be diagonalised to yield the mass matrices:
\begin{align}
&m_\text{light} \simeq  m_D m_R^{-1} m_D^T \in \mathbb{C}^{n_L\times n_L},\,\nonumber\\
&m_\text{heavy} \simeq m_R \in \mathbb{C}^{n_R\times n_R}.
\end{align}
The spectrum contains $(n_L-n_R)$ massless states at leading order. Therefore, the light neutrino mass matrix can be written as
\begin{equation}\label{ECIMassMatrixb}
    m_\nu= Y_1^T\, M_{11} \,Y_1\,,
\end{equation}
where $Y_1 = Y^T$ and $M_{11}=v^2/2\, m^{-1}_R$. Therefore, it has the form of Eq.~\eqref{completemassmatrix2} with $M_{12}=M_{21}=M_{22}=0$, and the usual CI parametrisation~\cite{Casas:2001sr} can be used to express the Yukawa $Y_1$ in terms of light neutrino observables and other unknown free parameters. The most general neutrino Yukawa coupling compatible with low energy data, in the basis where the charged leptons Yukawa couplings and the $M_{11}$ are diagonal, reads:
\begin{align}
Y_{1}&=D^{-1/2}_{M_{11}}\,R_{\ell}\,D_{\sqrt{m}}\,U^\dagger\,\nonumber\\
&=\frac{\sqrt{2}}{v}\,m_R^{1/2}\,R_{\ell}\,D_{\sqrt{m}}\,U^\dagger\,\\ &\text{Standard CI parametrisation},\nonumber
\end{align}
where $D^{-1/2}_{M_{11}}$ is the diagonal matrix of the square roots of the eigenvalues of $M_{11}$, $D_{\sqrt{m}}$ is the diagonal matrix of roots of the physical masses $m_i$ of the light neutrinos, and $R_{\ell}$ is the $n_R\times n_L$ complex orthogonal matrix that includes all the free parameters describing the neutrino sector. In Sec.~\ref{sec:CIgen}, we discuss a rewriting of the general neutrino mass matrix in Eq.~\eqref{eq:MMSS2}, which could potentially serve to obtain a GCI parametrisation outside of the seesaw limit, or one that includes higher order corrections. We also show in App.~\ref{app:oldCI} how the standard CI may be derived in the seesaw limit using this approach.

Several other models exist in the literature where only the $M_{11}$ element is different from zero at tree level, like Type-III seesaw~\cite{Foot:1988aq}, Model $A_1$ in Ref.~\cite{Giarnetti:2023osf}, and other variants of the Type-I seesaw~\cite{Branco:2020yvs,CentellesChulia:2024uzv,Grimus:2001ex}.

\subsubsection{Loop level: the Scotogenic Model} \label{sec:ScM}

The Scotogenic Model (ScM)~\cite{Tao96, Ma:2006km} is based on a $\mathbb{Z}_2$ symmetry, see also Ref.~\cite{Ma:1998dn}. The particle content with the associated charges can be found in Table~\ref{ScMparticles}. The relevant terms in the Lagrangian of the model are:
\begin{equation}\label{eq:lagrangianScM}
-\mathcal{L}_{\rm ScM} = \frac{1}{2} \overline{\psi_R^c} m_R\psi_R  + \left( \overline{\psi_R} \tilde{\Phi}^{\dagger}  YL + \text{H.c.} \right)\,,
\end{equation}
where $ \tilde{\Phi} \equiv i\sigma_2 \Phi^* $. As in the case of the seesaw, at least two copies of the fermion singlets are needed. The scalar potential for $\Phi$ and the SM Higgs doublet $H$ is given by:
\begin{align}
\label{potentialSc}
\mathcal{V}_{\rm ScM}  =& \,-\,m_H^2  H^\dagger H + \lambda_H (H^\dagger H)^2 \,
+ \, m_\Phi^2  \Phi^\dagger\Phi\, +\, \lambda_\Phi (\Phi^\dagger\Phi )^2\,
\nonumber\\&+ \, \lambda_{H\Phi} (H^\dagger H) (\Phi^\dagger\Phi)  
+ \lambda_{H\Phi,2} (H^\dagger \Phi) (\Phi^\dagger H)   
\,\nonumber\\
&+\, \frac{\lambda_{H\Phi,3}}{2}\left[ (H^\dagger \Phi )^2 \, +\, {\rm H.c.} \right]
\,. 
\end{align}
The scalar mass spectrum, derived from the potential in Eq.~\eqref{potentialSc}, reads
\begin{equation}
\begin{aligned}
	m_{\phi_0^R} & =  \sqrt{m_\Phi^2\,+\,\frac 1 2 \left(\lambda_{H\Phi}\,+\,\lambda_{H\Phi,2}\,+\,\lambda_{H\Phi,3} \right) v^2} \,,\\
	m_{\phi_0^I} &=\sqrt{m_\Phi^2\,+\,\frac 1 2 \left(\lambda_{H\Phi}\,+\,\lambda_{H\Phi,2}\,-\,\lambda_{H\Phi,3} \right) v^2}\,,\\
	m_{\phi^+} &= \sqrt{m_{\Phi}^2\,+\,\frac 1 2 \lambda_{H\Phi}  \,v^2} \,.
\end{aligned}
\end{equation}
Here $\phi_0^R$ and $\phi_0^I$ denote the real and imaginary parts of the neutral field $\phi_0$ in the additional scalar doublet, and $\phi^+$ its charged component. The scalars $\phi_0^R$ and $\phi_0^I$ acquire a mass splitting proportional to the quartic coupling $\lambda_{H\Phi,3}$; taking $\lambda_{H\Phi,3}$ small is technically natural in the 't~Hooft sense~\cite{tHooft:1979rat}.

 \renewcommand{\arraystretch}{1.5}
\begin{table}[!t]
    \centering
    \begin{tabular}{l c c c c}
        \hline
        \hline
        Field & SU(3)$_C$ & SU(2)$_L$ & U(1)$_Y$ & $\mathbb{Z}_2$ \\
        \hline
        $\Phi \equiv (\phi^+, \phi_0)^T$ & 1 & 2 & $\phantom{-}1/2$ & 1 \\
        $\psi_R$ & 1 & 1 & $\phantom{-}0$ & 1 \\
        \hline
    \end{tabular}
    \caption{Particle content and charge assignments of the Scotogenic Model.}
    \label{ScMparticles}
\end{table}
\renewcommand{\arraystretch}{1}

\begin{figure}[t]
	\begin{tikzpicture}[node distance=1cm and 1cm]
     \coordinate[label=left:$L$] (nu1);
     \coordinate[vertex, right=of nu1] (v1);
     \coordinate[cross, right=of v1] (lfv);
     \coordinate[vertex, above=of lfv] (v3);
     \coordinate[above left=of v3,  label=left:$H$] (h1);
     \coordinate[above right=of v3,  label=right:$H$] (h2);
     \coordinate[vertex, right=of lfv] (v2);
     \coordinate[right=of v2, label=right:$L$] (nu2);
     \coordinate[above=of v1, xshift=0.25cm, yshift=-0.3cm, label=left:$\Phi$] (s1);
     \coordinate[above=of v2, xshift=-0.25cm, yshift=-0.3cm, label=right:$\Phi$] (s2);

     \draw[fermion] (nu1)--(v1);
     \draw[fermion] (v1) -- node[below]{$\psi_{R}$} ++ (lfv);
     \draw[fermion] (v2)-- node[below]{$\psi_{R}$} ++ (lfv);
     \draw[fermion] (nu2)--(v2);
     \draw[scalar] (h1) -- (v3);
     \draw[scalar] (h2) -- (v3);
     \draw[scalar] (v3) to[out=180,in=90] (v1);
     \draw[scalar] (v3) to[out=0,in=90] (v2);
   \end{tikzpicture}
%
\caption{Neutrino mass contributions at the one-loop level in the Scotogenic Model. The neutral scalars $\phi^R_0$ and $\phi_0^I$ run in the loop.
\label{fig:ScM}}
\end{figure}
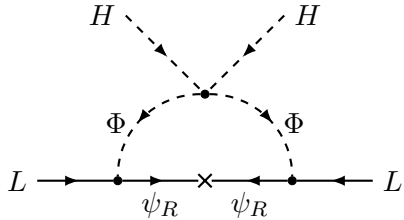

The exact discrete $\mathbb{Z}_2$ prevents the singlet fermion to couple with the Higgs doublet and neutrinos remain massless at tree level. The latter only obtain a mass at one loop from the diagram shown in Fig.~\ref{fig:ScM}. The Majorana neutrino mass matrix reads:
\begin{equation}\label{ScM}
   (m_\nu)_{\alpha \beta}=\dfrac{1}{32\pi^2} \sum_i \left(Y_{\alpha i} Y_{\beta i} m_{R_i}\,\right)  \mathcal{F}(m_{\phi^R_0},m_{\phi_0^I},m_{R_i})\,,
\end{equation}
where the loop function is
\begin{equation} \label{eq:Floop}
\mathcal{F}(x,y,z)=\dfrac{x^2}{x^2-z^2}\log \dfrac{x^2}{z^2}-\dfrac{y^2}{y^2-z^2}\log \dfrac{y^2}{z^2}\,.
\end{equation}
The neutrino mass matrix has the form of Eq.~\eqref{ECIMassMatrixb}, $m_\nu= Y_1^T\, M_{11} \,Y_1$ with
\begin{equation}
    Y_{1}=Y^T\,,\qquad M_{11}= \dfrac{m_{R_i}}{32\pi^2}\, \mathcal{F}(m_{\phi^R_0},m_{\phi_0^I},m_{R_i})\,.
\end{equation}
Therefore, the seesaw-like flavour structure allows to use the standard CI parametrisation~\cite{Casas:2001sr}, after trivially taking into account the different global factor.
This result is in agreement with that found in Ref.~\cite{Ibarra:2016dlb}, where the ScM was studied in the context of direct detection of Dark Matter.

At one loop, another similar example is the ScotoSinglet Model~\cite {Beniwal:2020hjc}; see Refs.~\cite{Cacciapaglia:2020psm,Bonnet:2012kz,Restrepo:2013aga} for other one-loop models and Ref.~\cite{Cai:2017jrq} for a review of radiative neutrino mass models.

\subsection{Off-diagonal mass matrix}

In the following we consider the scenario where only the crossed terms are non-vanishing, i.e., $M_{11}=M_{22}=0$. We want to show how the CI parametrisation can be implemented for such a broad type of scenario, without considering a particular model; this means that we do not consider if the neutrino mass matrix leads to a massless neutrino eigenvalue or to a specific mass ordering. We will go through some other examples in the following sections. The light neutrino mass matrix reads of Eq.~\eqref{completemassmatrix2} reads
\begin{equation} \label{eq:off}
m_\nu=Y_1^TM_{12}Y_2+Y_2^TM_{12}^TY_1\,,
\end{equation}
which can be rewritten in the from of Eq.~\eqref{ECIMassMatrixb} with
\begin{equation}\label{eq:1case}
    \mathcal{Y}=\begin{pmatrix}
        Y_1\\
        Y_2
    \end{pmatrix}\, \qquad \text{and}\quad\,\mathcal{M}=\begin{pmatrix}
        0&M_{12}\\
        M_{12}^T&0
    \end{pmatrix}\,.
\end{equation}
The extended Yukawa matrix is $(n_1+n_2)\times n_L$, and the extended mass matrix $\mathcal{M}$ is $(n_1+n_2)\times (n_1 + n_2)$.
The matrix $\mathcal{M}$ can be diagonalised by  
\begin{equation} \label{eq:Vgen}
    V=\dfrac{1}{\sqrt{2}}\,\begin{pmatrix}
        i\,\mathbb{I}_{n_1\times n_1}&-i\,\mathbb{I}_{n_1\times n_2}\\
        \mathbb{I}_{n_2\times n_1}&\mathbb{I}_{n_2\times n_2}
    \end{pmatrix}\,,
\end{equation}
for arbitrary $n_1$ and $n_2$, where the presence of the $i$ ensures that all 
eigenvalues are positive, and with $D_\mathcal{M} = {\rm diag}(M_{12},M_{12})$. Therefore, we can proceed with the GCI parametrisation provided in Eq.~\eqref{eq:CIgen}. 
\vspace{2em}
\\
{\bf{Case $\mathbf{n_1=n_2=1}$}\vspace{.3cm}\\}
In this case, the Yukawa couplings are three-dimensional vectors, so that the rank of the neutrino mass matrix is two. This implies that such a mass structure provides one massless neutrino. There is one parameter stemming from the $\mathcal{M}_{2\times2}$ matrix, together with the 12 real parameters needed to specify the Yukawa interactions. As is well known, not all of them are physical quantities. Indeed, three of them can be absorbed by a rotation of the lepton doublets and another one by the rescaling of the parameter in $\mathcal{M}_{2\times2}$. Therefore, there are only 10 effective parameters that determine the light-neutrino sector. Finally, we observe that the measurable parameters yield 2 masses, 3 mixing angles, and 2 phases, for a total of 7 measurables, and we conclude that in order to describe the Majorana neutrino mass with off-diagonal mass matrix and $n_1=n_2=1$, we need two real parameters. These parameters are included in the matrix $\mathcal{R}$, which can be obtained by following the procedure outlined in Eqs.~\eqref{general3x3R} and ~\eqref{eq:cut3x3}, setting $\theta_{13} = \theta_{23} = 0$:\footnote{Since we need only one complex parameter two of the angles can be put to zero. The choice of which angles to set to zero depends on the structure of the diagonal neutrino mass matrix, i.e., on the position of its non-vanishing eigenvalues. By choosing the two non-zero eigenvalues to lie in the $2\times2$ upper-left sub-block of $D_m$, only $\theta_{12}$ has to appear in $\mathcal{R}$.}
\begin{equation}\label{R2x3}
    \mathcal{R}_{2\times 3}=\begin{pmatrix}
        \cos z&\mp \sin z&0\\
        \sin z&\pm \cos z&0
    \end{pmatrix}\,,
\end{equation}
where $z$ is a complex angle, yielding the 2 real parameters discussed above. It must be noted that a $\pm$ in the second column has been included to account for possible reflections of the orthogonal $\mathcal{R}$ matrix.
Let us now discuss how to diagonalise the $\mathcal{M}$ matrix in Eq.~\eqref{eq:1case}. Using Eq.~\eqref{eq:Vgen}, the matrix $V$ of Eq.~\eqref{eq:Vdiag} can be directly obtained.

In standard seesaw models, the mixing of the charged leptons typically scales as $\Theta\sim \sqrt{m_\nu/M_N}$, with $M_N$ the mass of the heavy leptons that leads to a suppression of the size of the Yukawa couplings,
unless special mass textures are imposed in accordance with lepton number conservation~\cite{WYLER1983205,Mohapatra:1986aw,Mohapatra:1986bd,BERNABEU1987303,Branco:1988ex,Barr:2003nn,Kersten:2007vk,Gavela:2009cd,Hernandez:2022ivz}. 
This feature is shared by both the Linear seesaw and the Generalised-Scotogenic Models, discussed below as examples of tree-level and loop-level neutrino mass generation, respectively. 
In the Linear seesaw scenario, lepton number violation (LNV) arises only through a small perturbation in the mass matrix, while in the GScM non-zero tree level neutrino masses are protected by an additional $U(1)_\text{DM}$ symmetry. 
As a consequence, in both models the violation of lepton number is strongly suppressed, which naturally leads to a hierarchical Yukawa structure, i.e., $Y_2 \ll Y_1$. 
Therefore, within the CI parametrisation, it is immediate to take the limit of large imaginary parts of the complex angle $z$,\footnote{In the large imaginary part limit of the angle $z$, the imaginary component is assumed to be positive.
} corresponding to this strong hierarchy, which yields a rich phenomenology.
For Inverted Ordering (IO), we obtain:
\begin{align} \label{eq:simple_parIO}
    &{Y_1}=\dfrac{e^{-\text{Im}z}}{\sqrt{2 \overline{M}_1}}\Big[ u_1^\ast \sqrt{m_1}\mp i\,u_2^\ast\sqrt{m_2}\Big] \,,\nonumber\\
    &Y_2=\dfrac{e^{\text{Im}z}}{\sqrt{2 \overline{M}_1}}\Big[ u_1^\ast \sqrt{m_1}\pm i\,u_2^\ast\sqrt{m_2}\Big]\,,
\end{align}
while for the Normal Ordering (NO), we have:
\begin{align} \label{eq:simple_parNO}
    &{Y_1}=\dfrac{e^{-\text{Im}z}}{\sqrt{2 \overline{M}_1}}\Big[ u_2^\ast \sqrt{m_2}\mp i\,u_3^\ast\sqrt{m_3}\Big] \,,\nonumber\\
    &Y_2=\dfrac{e^{\text{Im}z}}{\sqrt{2 \overline{M}_1}}\Big[ u_2^\ast \sqrt{m_2}\pm i\,u_3^\ast\sqrt{m_3}\Big]\,,
\end{align}
where $u_i$ are the two columns of the PMNS mixing matrix.
It is worth to notice that this result is the large mixing angle limit of the parametrisation of Ref.~\cite{Ibarra:2003up}, obtained for the seesaw scenario with two right-handed neutrinos.
Also, a similar parametrisation can be found in Ref.~\cite{Ibarra:2011xn} in the context of low scale seesaw with two right-handed neutrinos.
\\[2em]
{\bf{Case $\mathbf{n_1=n_2=2}$}\vspace{.3cm}\\}
The matrix $V$ which diagonalises the matrix $\mathcal{M}_{4\times 4}$ to the diagonal form 
\begin{equation}
    D_\mathcal{M}=\text{diag}(\overline{M}_1,\overline{M}_2,\overline{M}_1,\overline{M}_2)\,,
\end{equation}
through the Autonne-Takagi decomposition in Eq.~\eqref{eq:Vdiag} is given by Eq.~\eqref{eq:Vgen}. Moreover, due to the increased dimensions of the Yukawa coupling matrices and of the matrix $\mathcal{M}_{4\times 4}$, there are more free parameters to consider. With a similar parameter counting as the one performed for the previous case, we find that in this scenario there are 12 free real parameters. Using Eq.~\eqref{eq:Rnm_cases}, we parametrise $\mathcal{R}_{4\times3}$ as:
\begin{equation}
    \mathcal{R}_{4\times 3}=\mathcal{R}_{\,4\times 4} \,
\begin{pmatrix}
\mathbb{I}_{3} \\ \hline
0_{1\times 3}
\end{pmatrix}\,,
\end{equation}
which depends on 6 complex angles, contributing to the 12 real parameters required to describe the neutrino sector in this scenario.\\[2em]
{\bf{Case $\mathbf{n_1=n_2=3}$}\vspace{.3cm}\\}
Using Eq.~\eqref{eq:Vgen}, the matrix $V$ which diagonalises the matrix $\mathcal{M}_{6 \times 6}$ to the diagonal form
\begin{equation*}
    D_\mathcal{M}=\text{diag}(\overline{M}_1,\overline{M}_2,\overline{M}_3,\overline{M}_1,\overline{M}_2,\overline{M}_3)
\end{equation*}
through the Autonne-Takagi decomposition in Eq.~\eqref{eq:Vdiag} can be directly obtained. With a similar parameter counting performed as for the previous cases, we find that in this scenario there are 24 free real parameters. Using Eq.~\eqref{eq:Rnm_cases}, we parametrise $\mathcal{R}_{6\times3}$ as:
\begin{equation}
    \mathcal{R}_{6\times 3}=\mathcal{R}_{\,6\times 6} \,
\begin{pmatrix}
\mathbb{I}_{3} \\ \hline
0_{3\times 3}
\end{pmatrix}\,,
\end{equation}
which, in general, depends on 15 angles, for a total of 30 real parameters. 
Similarly to the case with $n_1 = n_2 = 1$, since there are more free parameters than needed to describe the neutrino sector, some of the angles can be set to zero. 
In this scenario, 3 angles turn out to be redundant.
\begin{figure}[tb]
\centering
\begin{tikzpicture}[node distance=1.5cm and 1.cm]

  \coordinate[label=left:$L$] (L1);
  \coordinate[vertex, right=of L1] (v1);
  \coordinate[cross, right=of v1] (massX);
  \coordinate[vertex, right=of massX] (v2);
  \coordinate[right=of v2, label=right:$L$] (L2);

  \coordinate[above=of v1, label=above:$H$] (H1);
  \coordinate[above=of v2, label=above:$H$] (H2);

  \draw[fermion] (L1) -- node[below]{} (v1);
  \draw[fermion] (v1) -- node[below]{$N$} (massX);
  \draw[fermion] (massX) -- node[below]{$S$} (v2);
  \draw[fermion] (L2) -- node[below]{} (v2);

  \draw[scalar] (H1) -- (v1);
  \draw[scalar] (H2) -- (v2);

  \node[above=2pt of massX] {};
\end{tikzpicture}
\begin{tikzpicture}[node distance=1.5cm and 1.cm]

  \coordinate[label=left:$L$] (L1);
  \coordinate[vertex, right=of L1] (v1);
  \coordinate[cross, right=of v1] (massX);
  \coordinate[vertex, right=of massX, label=above:$\mu$] (v3);
  \coordinate[cross, right=of v3] (v4);
  \coordinate[vertex, right=of v4] (v2);
  \coordinate[right=of v2, label=right:$L$] (L2);

  \coordinate[above=of v1, label=above:$H$] (H1);
  \coordinate[above=of v2, label=above:$H$] (H2);

  \draw[fermion] (L1) -- node[below]{} (v1);
  \draw[fermion] (v1) -- node[below]{$N$} (massX);
  \draw[fermion] (massX) -- node[below]{$S$} (v3);
  \draw[fermion] (v4) -- node[below]{$S$} (v3);
  \draw[fermion] (v2) -- node[below]{$N$} (v4);
  \draw[fermion] (L2) -- node[below]{} (v2);

  \draw[scalar] (H1) -- (v1);
  \draw[scalar] (H2) -- (v2);

  \node[above=2pt of massX] {};
\end{tikzpicture}
\caption{Neutrino masses at tree level in the Linear seesaw (left) and the Inverse seesaw (right).} \label{fig:LSSfeynman}
\end{figure}
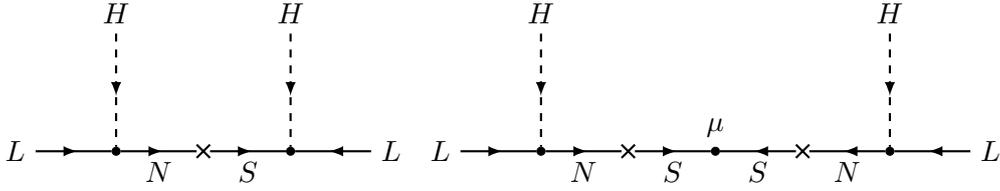
\subsubsection{Tree level: the Linear Seesaw Model}
The Linear seesaw (LSS) was first introduced in Refs.~\cite{Akhmedov:1995ip,Akhmedov:1995vm} for left-right symmetric models and also in grand unified theories~\cite{Barr:2003nn,Malinsky:2005bi}. Variations of the neutrino mass matrix have been proposed in Refs.~\cite{Abada:2007ux, Gavela:2009cd}. More recently, models with new scalars in higher $SU(2)$ representations have been proposed~\cite{Giarnetti:2023osf,Giarnetti:2023dcr}. In general, the model consists of $n_1$ right-handed neutrinos $N$ and $n_2$ singlet fermions $S$, so that the relevant terms in the Lagrangian are:
\begin{equation} \label{eq:LSS}
    -\mathcal{L}_\text{LSS}= \tilde H^\dagger\overline{N}Y_N L+\overline{\tilde L} H Y_S S+ \overline{N}\mathrm{M}_RS+\text{ H.c.}\,,
\end{equation}
where $ \tilde{L} \equiv i\sigma_2 C \bar{L}^T $, $ C $ the charge conjugation matrix, and $\mathrm{M}_R$ is a $n_1\times n_2$ complex mass matrix and $Y_N$ and $Y_S$ are $3\times n_1$ and $3\times n_2$ complex Yukawa matrices, respectively. In the basis $(\nu_L, N^c,S)$, after the EW SSB, the $(n_1+n_2+3)\times(n_1+n_2+3)$ neutral fermion mass matrix reads
\begin{equation}
    M_\nu=\begin{pmatrix}
        0&\frac{v}{\sqrt{2}}\,Y_N^T&\frac{v}{\sqrt{2}}\,Y_S\\
        \frac{v}{\sqrt{2}}\,Y_N&0&\mathrm{M}_R\\
        \frac{v}{\sqrt{2}}\,Y_S^T&\mathrm{M}_R^T&0
    \end{pmatrix}\,.
\end{equation}
\begin{table}[!htb]
    \centering
    \begin{tabular}{l c c c c}
        \hline
        \hline
        Field & SU(3)$_C$ & SU(2)$_L$ & U(1)$_Y$ & U(1)$_{\text{DM}}$ \\
        \hline
        $\Phi \equiv (\phi^+, \phi_0)^T$ & 1 & 2 & $\phantom{-}1/2$ & 1 \\
        $\Phi' \equiv ({\phi_0}', {\phi^-}')^T$ & 1 & 2 & $-1/2$ & 1 \\
        Vector-like fermion $\psi$ & 1 & 1 & $\phantom{-}0$ & 1 \\
        \hline
    \end{tabular}
    \caption{Particle content and charge assignments of the Generalised Scotogenic Model.}
    \label{GScMparticles}
\end{table}
Taking the limit $Y_N,Y_S\ll \mathrm{M}_R/v$, the mass matrix of the light neutrinos is given by
\begin{equation}\label{eq:LSSmnu}
    m_\nu=\dfrac{v^2}{2}\left( Y_S \mathrm{M}_R^{-1}Y_N+Y_N^T{\mathrm{M}_R^{T}}^{-1}Y_S^T\right)\,,
\end{equation}
which has the form of Eq.~\eqref{eq:off} with
\begin{equation}
    Y_{1}=Y_S^T\,,\qquad Y_{2}= Y_N\,,\qquad M_{12}=\frac{v^2}{2} \mathrm{M}_R^{-1}\,.
\end{equation}
Note that the heavy fermions are degenerate in the aforementioned limit. The origin of the name \emph{linear seesaw} becomes evident: the Yukawa couplings of the right-handed neutrinos enter linearly in the mass matrix, in contrast to the usual seesaw in which they enter quadratically. 
A schematic diagram of the neutrino mass generation is shown in Fig.~\ref{fig:LSSfeynman}.

One may directly use the GCI parametrisation, Eq.~\eqref{eq:CIgen}, with $V$ given by Eq.~\eqref{eq:Vgen}. For the $n_1=n_2=1$ case, the simpler forms for the Yukawas provided in Eqs.~\eqref{eq:simple_parIO} and~\eqref{eq:simple_parNO} may be used for IO and NO, respectively. Note that $Y_S$ violates lepton number in two units and therefore it is small in 't Hooft sense~\cite{tHooft:1979rat}. This correspond to the large imaginary part of the complex angle $z$. This flavour structure also applies to Models $A_2$ and $B_i$ in Ref.~\cite{Giarnetti:2023osf}. The parametrisation presented for the Linear seesaw model can be mapped onto the parametrisation used in Ref.~\cite{Hernandez:2022ivz} in the case of degenerate heavy neutrinos by taking $\text{Re[z]}\rightarrow \theta/2$. Note that in that reference the Majorana phase is shifted by $\pi/2$ in order to provide all positive neutrino masses, whereas here the result is automatically obtained. 

\begin{figure}[!t]
\centering
  \centering
 	\begin{tikzpicture}[node distance=1cm and 1cm]
     \coordinate[label=left:$L$] (nu1);
     \coordinate[vertex, right=of nu1] (v1);
     \coordinate[cross, right=of v1] (lfv);
     \coordinate[vertex, above=of lfv] (v3);
     \coordinate[above left=of v3,  label=left:$H$] (h1);
     \coordinate[above right=of v3,  label=right:$H$] (h2);
     \coordinate[vertex, right=of lfv] (v2);
     \coordinate[right=of v2, label=right:$L$] (nu2);
     \coordinate[above=of v1, xshift=0.25cm, yshift=-0.3cm, label=left:$\Phi$] (s1);
     \coordinate[above=of v2, xshift=-0.25cm, yshift=-0.3cm, label=right:$\Phi'$] (s2);

     \draw[fermion] (nu1)--(v1);
     \draw[fermion] (v1) -- node[below]{$\psi$} ++ (lfv);
     \draw[fermion] (lfv)-- node[below]{$\psi$} ++ (v2);
     \draw[fermion] (nu2)--(v2);
     \draw[scalar] (h1) -- (v3);
     \draw[scalar] (h2) -- (v3);
     \draw[scalar] (v3) to[out=180,in=90] (v1);
     \draw[scalar] (v3) to[out=0,in=90] (v2);
   \end{tikzpicture}
%
\caption{Neutrino masses at the one-loop level in the Generalised Scotogenic Model.}
\label{fig:GScM}
\end{figure}
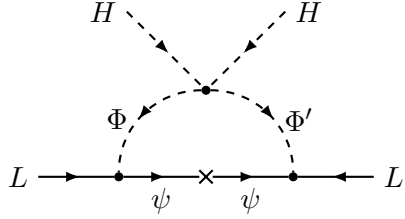
\subsubsection{Loop level: the Generalised Scotogenic Model}
\label{sec:GScM}
Here we discuss the Generalised Scotogenic Model (GScM)~\cite{Hagedorn:2018spx}, a generalisation of the ScM Model~\cite{Tao96, Ma:2006km} based on a global $ \text{U}(1)_{\text{DM}} $ symmetry instead of the $\mathbb{Z}_2$ symmetry of ScM. We follow closely the notation of Ref.~\cite{Hagedorn:2018spx}. The SM is augmented by two additional scalar doublets and at least one vector-like Dirac fermion, all charged under the $ \text{U}(1)_{\text{DM}} $ symmetry. The particle content with the associated charges can be found in Table~\ref{GScMparticles}. The relevant terms in the Lagrangian are
\begin{equation}\label{eq:lagrangianGScM}
-\mathcal{L}_{\rm GScM} = \overline{\psi} M_{\psi} \overline{\psi} + \left(\overline{\psi} Y\tilde{\Phi}^{\dagger} L + \overline{\psi} Y^{\prime*} \tilde{\Phi'}^{\dagger} \tilde{L} + \text{H.c.} \right)\,.
\end{equation}

For the case of just one fermion, the neutrino Yukawa couplings $Y, Y^\prime$ are three-component row vectors. The scalar potential of $\Phi,\,\Phi'$ and $H$ is given by
\begin{align}
\mathcal{V}_{\rm GScM} &= - m_H^2 H^\dagger H + \lambda_H (H^\dagger H)^2 + m_{\Phi}^2 \Phi^\dagger \Phi + \lambda_{\Phi} (\Phi^\dagger \Phi)^2\nonumber\\ 
&+ m_{\Phi'}^2 \Phi'^\dagger \Phi' + \lambda_{\Phi'} (\Phi'^\dagger \Phi')^2
+\lambda_{H\Phi} (H^\dagger H)(\Phi^\dagger \Phi)\nonumber\\
&+ + \lambda_{H\Phi'} (H^\dagger H)(\Phi'^\dagger \Phi')  +\lambda_{\Phi \Phi'} (\Phi^\dagger \Phi)(\Phi'^\dagger \Phi') 
\nonumber\\
&+ \lambda_{H\Phi,2} (H^\dagger \Phi)(\Phi^\dagger H) 
+ \lambda_{H\Phi',2} (H^\dagger \tilde{\Phi}')( \tilde{\Phi}'^\dagger H) \nonumber\\
&+ \lambda_{\Phi \Phi',2} (\Phi^\dagger \tilde{\Phi}')( \tilde{\Phi}'^\dagger \Phi)
+ \lambda_{H\Phi \Phi'} \left[ (H^\dagger \tilde{\Phi}') (H^\dagger \Phi) 
+ \text{H.c.} \right] .
\label{scalarpotential}
\end{align}

In the interaction basis $\left( H_0,\Phi_0,\Phi_0^\prime\right)$, the scalar mass matrix is:
\begin{widetext}
\begin{equation}
    M^2_S=\begin{pmatrix}
        -m_H^2+2v^2\lambda_H&0&0\\
        0&m^2_\Phi+\frac{1}{2}v^2\left(\lambda_{H\Phi}+\lambda_{H\Phi,2}\right)&-\frac{1}{2}v^2\lambda_{H\Phi\Phi^\prime}\\
        0&-\frac{1}{2}v^2\lambda_{H\Phi\Phi^\prime}&m^2_{\Phi^\prime}+\frac{1}{2}v^2\left(\lambda_{H\Phi^\prime}+\lambda_{H\Phi^\prime,2}\right)
    \end{pmatrix}\,.
\end{equation}
\end{widetext}
The neutral mass eigenstates are defined as:
\begin{equation}
    \begin{aligned}
        \eta_0=&\phantom{-}s_\theta \Phi_0+c_\theta \Phi_0^\prime\,,\\
        \eta_0^\prime=&-c_\theta \Phi_0+s_\theta \Phi_0^\prime\,,\\
    \end{aligned}
\end{equation}
with $c_\theta\equiv\cos\theta$ and $s_\theta\equiv\sin\theta$. The mixing angle is 
\begin{equation}
    \tan2\theta=\dfrac{2c}{b-a}\,,\label{eq:mixingangle}
\end{equation}
with
\begin{equation}
    \begin{aligned}
        c=&-\dfrac{1}{2}\lambda_{H\Phi\Phi^\prime}v^2\,,\\
        a=&m_\Phi^2+\dfrac{1}{2}v^2\left( \lambda_{H\Phi}+\lambda_{H\Phi,2}\right)\,,\\
        b=&m_{\Phi^\prime}^2+\dfrac{1}{2}v^2\left( \lambda_{H\Phi^\prime}+\lambda_{H\Phi^\prime,2}\right)\,.
    \end{aligned}
\end{equation}
The mass eigenstates $(h,\eta_0,\eta_0^\prime)$ have masses:
\begin{equation}
\begin{aligned}
    m_h=&\sqrt{2\lambda_H} v\,,\\
    m_{\eta_0}=&\sqrt{\dfrac{1}{2}\left(a+b+\sqrt{\left(a-b\right)^2-4c^2}\right)}\,,\\
    m_{\eta_0^\prime}=&\sqrt{\dfrac{1}{2}\left(a+b-\sqrt{\left(a-b\right)^2-4c^2}\right)}\,.
    \end{aligned}
\end{equation}

Similarly to the Scotogenic Model, the exact global $U(1)_\text{DM}$ prevents the singlet fermion to couple with the Higgs doublet and neutrinos remain massless at tree level. The latter only obtain a mass at one loop from the diagram shown in Fig.~\ref{fig:GScM}. The Majorana neutrino mass matrix reads:
\begin{align}\label{GScM}
   (m_\nu)_{\alpha \beta}&=\dfrac{\sin2\theta}{32\pi^2} \sum_i \left(Y_{\alpha i}Y^\prime_{\beta i} M_i\,+ Y^{\prime}_{\alpha i} Y_{\beta i} M_i\,\right) \nonumber\\ 
  &\times \mathcal{F}(m_{\eta_0},m_{\eta_0^\prime},M_i)\,,
\end{align}
where the loop function is defined in Eq.~\eqref{eq:Floop}. The sum over the index $i$ indicates the generalisation to the case with multiple fermions, each with mass $M_i$.
The neutrino mass matrix has the form of Eq.~\eqref{eq:off} with
\begin{equation}
    Y_{1}=Y,\, Y_{2}= Y^{\prime},\, M_{12}= \dfrac{\sin 2\theta}{32\pi^2}\, {\rm diag} (M_i \mathcal{F}(M_i))\,.
\end{equation}
For the $n_1=n_2=1$ case, Eqs.~\eqref{eq:simple_parIO} and~\eqref{eq:simple_parNO} may be used for IO and NO, respectively. As a consistency check, we note that the parametrisation derived from the GCI construction coincides with that employed in Refs.~\cite{Cai:2014kra,Hagedorn:2018spx} by replacing $\zeta\rightarrow e^{-\text{Im}z}$. For suitable lepton number charge assignments of the additional scalar and fermion fields, lepton number violation originates from the Yukawa coupling $Y$, which can thus be regarded as naturally small in the sense of ’t~Hooft naturalness. This corresponds to the large imaginary part of the complex angle $z$.

Let us also mention that one of the interesting features of the GScM is that the lightest neutral particle of the dark sector (the dark fermion $\psi$ or the lightest neutral dark scalar) may be a good Dark Matter (DM) candidate. 

\subsection{Diagonal plus off-diagonal mass matrix}
This case corresponds to relaxing the assumption of vanishing diagonal entries in the matrix $\mathcal{M}$, thus obtaining the most general neutrino mass matrix $m_\nu$, namely that of Eq.~\eqref{completemassmatrix}, for which the GCI parametrisation applies, Eq.~\eqref{eq:CIgen}. As in the previous case with purely off-diagonal contributions, we will consider specific models in which neutrino masses arise either from tree-level diagrams (the Linear plus Inverse seesaw scenario) or from one-loop diagrams (the Extended Scotogenic Model).

\subsubsection{Tree level: the Linear plus Inverse Seesaw Model}
\label{sec:LISS}

In this section we consider the Linear plus Inverse seesaw Model (LISS). It adds a small Majorana mass for $S$,
\begin{equation}\label{eq:lagrangianISS}
-\mathcal{L}_{\rm ISS} = \frac{1}{2}  \bar{S^c}\, \mu\,S + \text{H.c.} \,,
\end{equation}
to the Lagrangian of the LSS model, $\mathcal{L}_{\rm LSS}$ given in Eq.~\eqref{eq:LSS},\footnote{One can also add another one in the $M_\nu^{11}$ element, but it will only contribute to neutrino masses at one loop. See Ref.~\cite{Gavela:2009cd,Dev:2012sg,CentellesChulia:2020dfh} for other variants of the inverse seesaw.} so that:
\begin{equation}\label{eq:lagrangianLISS}
\mathcal{L}_{\rm LISS} = \mathcal{L}_{\rm LSS} +\mathcal{L}_{\rm ISS}  \,.
\end{equation}
Note that $\mu$ violates lepton number in two units and therefore is small in t' Hooft sense. The tree-level neutrino mass generation in the (LSS) ISS is shown on the left (right) panel of Fig.~\ref{fig:LSSfeynman}. The neutrino mass  matrix reads: 
\begin{equation}
    M_\nu=\begin{pmatrix}
        0&\frac{v}{\sqrt{2}}\,Y_N^T&\frac{v}{\sqrt{2}}\,Y_S\\
        \frac{v}{\sqrt{2}}\,Y_N&0&\mathrm{M}_R\\
        \frac{v}{\sqrt{2}}\,Y_S^T&\mathrm{M}_R^T&\mu
    \end{pmatrix}\,.
\end{equation}
Similarly to the usual seesaw limit, we can assume $Y_N\,v,Y_S\,v, \mu\ll \mathrm{M}_R$, so that the mass matrix of the light neutrinos to the lowest order reads:
\begin{equation}
      m_\nu=\dfrac{v^2}{2}\left( Y_S \mathrm{M}_R^{-1}Y_N+Y_N^T{\mathrm{M}_R^{T}}^{-1}Y_S^T +Y_N^T {\mathrm{M}_R^{T}}^{-1}\mu {\mathrm{M}_R^{-1}} Y_N\right)\,,
\end{equation}
where the first (second) contribution corresponds to the linear (inverse) seesaw. This has the form of Eq.~\eqref{completemassmatrix2} with
\begin{align}
    &Y_{1}=Y_S^T,\, Y_{2}= Y_N,\,\nonumber\\
    &M_{22}=\frac{v^2}{2} {\mathrm{M}_R^{T}}^{-1}\, \mu \, {\mathrm{M}_R^{-1}},\,M_{12}=\frac{v^2}{2} \mathrm{M}_R^{-1}\,.
\end{align}
We now illustrate how the Yukawa parametrisation is modified in the presence of a non-zero diagonal mass term for the case $n_1=n_2=1$. In such a scenario, the $\mathcal{R}$ is the same as in Eq.~\eqref{R2x3} and $V$ is obtained from Eq.~\eqref{eq:Vgen}. This scenario differs from the LSS one due to the small contribution to the $\mathcal{M}$ eigenvalues proportional to the LNV term $\mu$. Therefore, now the heavy fermions form pseudo-Dirac pairs. The diagonalisation of $\mathcal{M}$ gives:
\begin{equation}
    V^\ast \begin{pmatrix}
        0&M_{12}\\
        M_{12}&M_{22}
    \end{pmatrix}V^\dagger\approx\begin{pmatrix}
        M_{12}+\frac{M_{22}}{2}&0\\
        0&M_{12}-\frac{M_{22}}{2}
    \end{pmatrix}\,,
\end{equation}
where $\approx$  indicates that this expression is valid in the limit $M_{22}\ll M_{12}$. 
For Inverted Ordering (IO), we obtain:
\begin{align} \label{eq:simple_parIOLSSIS}
    {Y_1}=&\dfrac{e^{-\text{Im}z}}{\sqrt{2 M_{12}}}\Big[ u_1^\ast \sqrt{m_1}\mp i\,u_2^\ast\sqrt{m_2}\Big] \nonumber\\
    +& \dfrac{e^{\text{Im}z}}{\sqrt{2 M_{12}}}\dfrac{M_{22}}{4M_{12}}\Big[u_1^\ast \sqrt{m_1}\pm i\,u_2^\ast\sqrt{m_2}\Big] \,,\nonumber\\
    Y_2=&\dfrac{e^{\text{Im}z}}{\sqrt{2 M_{12}}}\Big[ u_1^\ast \sqrt{m_1}\pm i\,u_2^\ast\sqrt{m_2}\Big]\,,
\end{align}
while for the Normal Ordering (NO), we have:
\begin{align} \label{eq:simple_parNOLSSIS}
    {Y_1}=&\dfrac{e^{-\text{Im}z}}{\sqrt{2 M_{12}}}\Big[ u_2^\ast \sqrt{m_2}\mp i\,u_3^\ast\sqrt{m_3}\Big]\nonumber\\
    +&\dfrac{e^{\text{Im}z}}{\sqrt{2 M_{12}}}\dfrac{M_{22}}{4M_{12}}\Big[ u_2^\ast \sqrt{m_2}\pm i\,u_3^\ast\sqrt{m_3}\Big] \,,\nonumber\\
    Y_2=&\dfrac{e^{\text{Im}z}}{\sqrt{2 M_{12}}}\Big[ u_2^\ast \sqrt{m_2}\pm i\,u_3^\ast\sqrt{m_3}\Big]\,,
\end{align}
where $u_i$ are the two columns of the PMNS mixing matrix. For completeness, we remark that also the $Y_2$ Yukawa receive contributions proportional to the LNV term; however, these are further suppressed by a factor $e^{-\text{Imz}}$ and are thus neglected.
As a consistency check, we find that our results reproduce those of Ref.~\cite{Hernandez:2022ivz} upon identifying $M_{22}\rightarrow \Delta M$ and $M_{12}\rightarrow M$.

\subsubsection{Loop level: the Extended Scotogenic Model}
\label{sec:ScMGSm}

 \renewcommand{\arraystretch}{1.5}
\begin{table}[!t]
    \centering
    \begin{tabular}{l c c c c}
        \hline
        \hline
        Field & SU(3)$_C$ & SU(2)$_L$ & U(1)$_Y$ & $\mathbb{Z}_2$ \\
        \hline
        $\Phi \equiv (\phi^+, \phi_0)^T$ & 1 & 2 & $\phantom{-}1/2$ & 1 \\
        $\psi_L$ & 1 & 1 & $\phantom{-}0$ & 1 \\
        $\psi_R$ & 1 & 1 & $\phantom{-}0$ & 1 \\
        \hline
    \end{tabular}
    \caption{Particle content and charge assignments of the Extended Scotogenic Model.}
    \label{EScMparticles}
\end{table}
\renewcommand{\arraystretch}{1}

\begin{figure}[tb]
\centering
	\begin{tikzpicture}[node distance=1cm and 1cm]
     \coordinate[label=left:$L$] (nu1);
     \coordinate[vertex, right=of nu1] (v1);
     \coordinate[cross, right=of v1] (lfv);
     \coordinate[vertex, above=of lfv] (v3);
     \coordinate[above left=of v3,  label=left:$H$] (h1);
     \coordinate[above right=of v3,  label=right:$H$] (h2);
     \coordinate[vertex, right=of lfv] (v2);
     \coordinate[right=of v2, label=right:$L$] (nu2);
     \coordinate[above=of v1, xshift=0.25cm, yshift=-0.3cm, label=left:$\Phi$] (s1);
     \coordinate[above=of v2, xshift=-0.25cm, yshift=-0.3cm, label=right:$\Phi$] (s2);

     \draw[fermion] (nu1)--(v1);
     \draw[fermion] (v1) -- node[below]{$\psi_{L\,(R)}$} ++ (lfv);
     \draw[fermion] (v2)-- node[below]{$\psi_{L\,(R)}$} ++ (lfv);
     \draw[fermion] (nu2)--(v2);
     \draw[scalar] (h1) -- (v3);
     \draw[scalar] (h2) -- (v3);
     \draw[scalar] (v3) to[out=180,in=90] (v1);
     \draw[scalar] (v3) to[out=0,in=90] (v2);
   \end{tikzpicture}
 \qquad
  \centering
 	\begin{tikzpicture}[node distance=1cm and 1cm]
     \coordinate[label=left:$L$] (nu1);
     \coordinate[vertex, right=of nu1] (v1);
     \coordinate[cross, right=of v1] (lfv);
     \coordinate[vertex, above=of lfv] (v3);
     \coordinate[above left=of v3,  label=left:$H$] (h1);
     \coordinate[above right=of v3,  label=right:$H$] (h2);
     \coordinate[vertex, right=of lfv] (v2);
     \coordinate[right=of v2, label=right:$L$] (nu2);
     \coordinate[above=of v1, xshift=0.25cm, yshift=-0.3cm, label=left:$\Phi$] (s1);
     \coordinate[above=of v2, xshift=-0.25cm, yshift=-0.3cm, label=right:$\Phi$] (s2);

     \draw[fermion] (nu1)--(v1);
     \draw[fermion] (v1) -- node[below]{$\psi$} ++ (lfv);
     \draw[fermion] (lfv)-- node[below]{$\psi$} ++ (v2);
     \draw[fermion] (nu2)--(v2);
     \draw[scalar] (h1) -- (v3);
     \draw[scalar] (h2) -- (v3);
     \draw[scalar] (v3) to[out=180,in=90] (v1);
     \draw[scalar] (v3) to[out=0,in=90] (v2);
   \end{tikzpicture}
%
\caption{Neutrino mass contributions at the one-loop level in the Extended Scotogenic Model. \emph{Left:} Contributions proportional to $m_{L\,(R)}$; \emph{Right:} Contributions proportional to $m_D$.}
\label{fig:ScSM}
\end{figure}
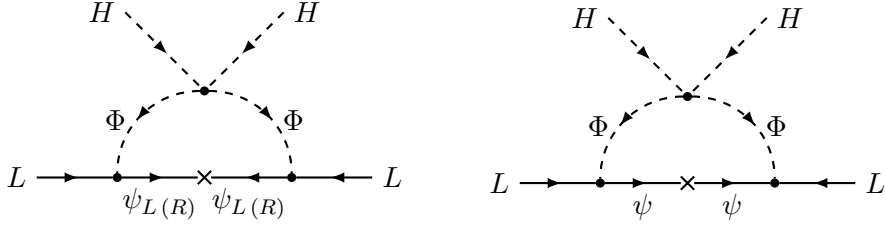
\noindent
Much like in the periodic table, where the presence of \emph{gaps} has guided the search for missing elements, the absence of a radiative model incorporating both diagonal and off-diagonal mass contributions motivates the following proposal. The model is the generalisation of the Scotogenic Model (ScM)~\cite{Tao96, Ma:2006km}, still based on a $\mathbb{Z}_2$ symmetry, but adding both chiralities of the fermion singlet. We term it the Extended Scotogenic Model (EScM). The particle content with the associated charges can be found in Table~\ref{EScMparticles}. The Lagrangian for the model is:
\begin{align}\label{eq:lagrangianEScM}
-\mathcal{L}_{\rm EScM} &= \bar{\psi} m_{D} \psi +\frac{1}{2} \overline{\psi_L^c} m_L\psi_L + \frac{1}{2} \overline{\psi_R^c} m_R\psi_R \nonumber\\
&+ \left( \overline{\psi_R} \tilde{\Phi}^{\dagger}  YL + \overline{\psi_L} \tilde{\Phi}^T Y^{\prime *} \tilde{L} + \text{H.c.} \right),
\end{align}
where for the case of just one copy of the fermion $\psi$, the neutrino Yukawa couplings $Y, Y^\prime$ are three-component row vectors. The scalar potential is provided in Eq.~\eqref{potentialSc}. In the basis $(\nu_L, \psi_R^c,\psi_L)$, the complete neutral fermion mass matrix at tree-level reads: 
\begin{equation}
    M^{\rm tree}_\nu=\begin{pmatrix}
        0&0&0\\
        0&m_R&m_D\\
        0&m^T_D&m_L
    \end{pmatrix}\,.
\end{equation}
Therefore, active neutrinos are massless at tree level. One can diagonalise the heavy fermion mass matrix. For one family of fermions, so that $m_R$, $m_D$ and $m_L$ are numbers, one gets:
\begin{align}\label{eq:egsScM}
M_{1,2} &= \frac{1}{2} \left[ 
m_{\rm R} + m_{\rm L}\pm\sqrt{(m_{\rm L} - m_{\rm R})^{2} + 4 m_{\rm D}^{2}})
\right] \,.
\end{align}
The neutral mass eigenstates are defined as:
\begin{equation}
    \begin{pmatrix}
        \psi_1 \\[4pt]
        \psi_2
    \end{pmatrix}
    =
    V(\theta)
    \begin{pmatrix}
        \psi^c_R \\[4pt]
        \psi_L
    \end{pmatrix},
    \qquad
    V(\theta) =
    \begin{pmatrix} \label{eq:V}
        \sin\theta & \cos\theta \\[4pt]
        -\cos\theta & \sin\theta
    \end{pmatrix}\,,
\end{equation}
with the mixing angle defined by
\begin{equation}
    \tan2\theta=\dfrac{2 m_D}{m_L-m_R}\,.\label{eq:mix}
\end{equation}
We can distinguish several limiting cases:
\begin{itemize}
  \item \textbf{Seesaw case:} 
  $m_{\rm R} \gg m_{\rm D} \gg m_{\rm L}$. We obtain:
  \begin{equation}
  M_1 \simeq m_L -\frac{m_D^2}{m_R}\,,\qquad\, M_2 \simeq m_R\,.
  \end{equation}
  Similarly for $m_{\rm L} \gg m_{\rm D} \gg m_{\rm R}$, interchanging $m_{\rm R} \leftrightarrow m_{\rm L}$. 
  \item \textbf{Pseudo-Dirac case:} $m_D \gg m_{L,R}$. We get:
  \begin{equation}
  M_{1} \simeq M_{2} \simeq m_{\rm D} \,.
  \end{equation}
  This scenario corresponds to the off-diagonal case discussed above.
\end{itemize}
Light neutrino masses are generated at one loop, as shown in the diagrams of Fig.~\ref{fig:ScSM}. The neutrino mass matrix can be written as in Eq.~\eqref{ECIMassMatrixa} with\footnote{Note that the contributions to the neutrino mass matrix can be equivalently expressed in the form of Eq.~\eqref{completemassmatrix2}.}
\begin{align}\label{EScMgeneral}
    &\mathcal{Y}=\begin{pmatrix}
        Y\\
        Y^\prime
    \end{pmatrix}\,,\nonumber\\
    &\mathcal{M}=\frac{1}{32\pi^2}\begin{pmatrix}
        F_1 M_1 s_\theta^2+F_2M_2c_\theta^2&(F_1 M_1 - F_2 M_2)s_\theta c_\theta\\
        (F_1 M_1 - F_2 M_2)s_\theta c_\theta&F_1 M_1 c_\theta^2+F_2M_2s_\theta^2
    \end{pmatrix}\,,
\end{align}
where $s_\theta=\sin\theta$ and $c_\theta=\cos\theta$, with the mixing angle $\theta$ defined in Eq.~\eqref{eq:mix}. Here $F_k \equiv \mathcal{F} (m_{\phi_0^R},m_{\phi_0^I},M_k)$, where the loop function $\mathcal{F}$ is given in Eq.~\eqref{eq:Floop}, and $M_1$ and $M_2$  are the mass eigenstates in Eq.~\eqref{eq:egsScM}. Importantly, even for the case of only one family of heavy fermions, two masses for light neutrinos are generated. It is therefore possible to use the GCI parametrisation, see also Ref.~\cite{Avila:2019hhv}. First, let us consider the seesaw case for the heavy fermions, $m_R \gg m_{D}\gg m_L$, such that $s_\theta \simeq m_D/m_R$. The elements of $\mathcal{M}$ are given by:
\begin{align}
    &M_{11} \simeq \frac{m_R}{32\pi^2}F_2\,,\nonumber\\
    &M_{12}\simeq -\frac{m_D}{32\pi^2}F_2\,,\nonumber\\
    &M_{22}\simeq \frac{1}{32\pi^2}\left[m_L\,F_1+(F_2-F_1)\frac{m^2_D}{m_R}\right]\,.
\end{align}
In the limit $M_k \gg M_{\phi_{0}}$, the light neutrino mass is approximately given by
\begin{equation}
    m_\nu \simeq \frac{1}{32\pi^2}Y'^T\,F_1\,\left(m_L -\frac{m_D^2}{m_R}\right)\,Y'\,.
\end{equation}
In this limit, the conventional seesaw scenario is reproduced, allowing to directly use the GCI parametrisation.
On the other hand, expanding $\mathcal{M}$ in Eq.~\eqref{EScMgeneral} in the pseudo-Dirac limit, $m_D \gg m_{L,R}$, we have $s_\theta \simeq 1/\sqrt{2}$, and obtain:
\begin{align}
    M_{11} &\simeq \frac{F_1+F_2}{64\pi^2}\left[m_R +m_D\, \frac{F_1-F_2}{F_1+F_2}\right]\,,\nonumber\\
    M_{12} &\simeq \frac{F_1+F_2}{64\pi^2}\left[m_D +\frac{1}{2}(m_L+m_R)\,\frac{F_1-F_2}{F_1+F_2}\right]\,,\nonumber\\
    M_{22} &\simeq \frac{F_1+F_2}{64\pi^2}\left[m_L +m_D \,\frac{F_1-F_2}{F_1+F_2}\right]\,.
\end{align}
Furthermore, neglecting the second term in the squared bracket, which is further suppressed by $(F_1-F_2)/(F_1+F_2) \ll 1$ as $M_1 \sim M_2 \simeq m_D$, we get the neat form
\begin{equation}
    \mathcal{M}\simeq\frac{F_1+F_2}{64\pi^2}\begin{pmatrix}
        m_R&m_D\\
        m_D&m_L
    \end{pmatrix}\,.
\end{equation}
Note that the $M_{11}$ contribution coincides with the Scotogenic one, while the $M_{12}+ M_{21}$ contribution is the same as in the Generalised Scotogenic Model. Equivalently,
\begin{align}
    m_\nu&=\frac{1}{32\pi^2} (F_1+F_2)[Y^T\,m_R\,Y + Y^{\prime T}\,m_L\,Y^\prime  \nonumber\\
    &+ Y^{T}\,m_D\,Y^\prime+Y^{\prime T}\,m_D\,Y]\,.
\end{align}
We thus recover the same neutrino mass structure discussed in the 
LSS+ISS case, with the difference being that in the present scenario $M_{11}\not=0$. The presence of this term modifies the form of the eigenvalues of $\mathcal{M}$, which now receive an additional contribution proportional to it, leading to an enhanced splitting between the eigenvalues of $\mathcal{M}$ (as discussed in Ref.~\cite{Hernandez:2022ivz} for the LSS + ISS case). 
The parametrisation of the Yukawa couplings, however, remains unchanged in its overall structure, and given by Eqs.~\eqref{eq:simple_parIOLSSIS} and~\eqref{eq:simple_parNOLSSIS} with $M_{22}$ now replaced by $M_{11} + M_{22}$.

\section{Implementation of the Generalised Casas-Ibarra parametrisation in models with extra symmetries} \label{sec:Zee}
The GCI parametrisation is particularly useful in scenarios characterised by anarchic Yukawa couplings, i.e., when there is no specific underlying symmetry in them. 
Nonetheless, the GCI framework remains applicable and insightful also in situations where the Yukawas have special symmetries. This is the case of an antisymmetric matrix, $f=-f^T$, which for instance appears in the case of a singly-charged singlet, $h^+$,
\begin{equation}
\overline{\tilde L}f L h^+ +\text{ H.c.}\,.
\end{equation}
As in this case neutrino masses cannot appear at tree level, we focus on a one-loop realisation, the Zee model.

\subsection{The Zee Model} 
\renewcommand{\arraystretch}{1.5}
\begin{table}[!t]
    \centering
    \begin{tabular}{l c c c}
        \hline
        \hline
        Field & SU(3)$_C$ & SU(2)$_L$ & U(1)$_Y$ \\
        \hline
        $H_1$& 1 & 2 & $1/2$ \\
        $H_2$& 1 & 2 & $1/2$ \\
        $h^+$ & 1 & 1 & $1$ \\
        \hline
    \end{tabular}
    \caption{Particle content and charge assignments of the Zee Model.}
    \label{Zeeparticles}
\end{table}
\renewcommand{\arraystretch}{1}
\noindent

As an illustrative example, we consider the Zee model~\cite{Zee:1980ai,Cheng:1980qt,Wolfenstein:1980sy}, see also Ref.~\cite{Herrero-Garcia:2017xdu} for a full analysis. The model provides an example of one-loop neutrino mass generation and consists of an extension of the Standard Model by an additional scalar doublet $H_2$, together with a singly-charged scalar singlet $h^+$. The particle content of the model is summarised in Table~\ref{Zeeparticles} and the neutrino mass generation at one-loop is shown in Fig.~\ref{fig:zee}. 
\begin{figure}[]
\centering
  \centering
 	\begin{tikzpicture}[node distance=1cm and 1cm]
     \coordinate[label=left:$L$] (nu1);
     \coordinate[vertex, right=of nu1] (v1);
     \coordinate[vertex, right=of v1] (lfv);
     \coordinate[vertex, above=of lfv] (v3);
     \coordinate[above=of v3,  label=left:$H_{1}$] (h1);
     \coordinate[below=of lfv,  label=left:$H_{1}$] (h2);
     \coordinate[vertex, right=of lfv] (v2);
     \coordinate[right=of v2, label=right:$L$] (nu2);
     \coordinate[above=of v1, xshift=0.25cm, yshift=-0.3cm, label=left:$h^-$] (s1);
     \coordinate[above=of v2, xshift=-0.25cm, yshift=-0.3cm, label=right:$H_{2}^-$] (s2);

     \draw[fermion] (nu1)--(v1);
     \draw[fermion] (lfv) -- node[below]{$L$} ++ (v1);
     \draw[fermion] (v2)-- node[below]{$e_R$} ++ (lfv);
     \draw[fermion] (nu2)--(v2);
     \draw[scalar] (h1) -- (v3);
     \draw[scalar] (h2) -- (lfv);
     \draw[scalar] (v1) to[out=90,in=180] (v3);
     \draw[scalar] (v3) to[out=0,in=90] (v2);
   \end{tikzpicture}
\caption{Neutrino masses at the one-loop level in the Zee Model. The mass eigenstates $h^+_{1,2}$ and the charged leptons run in the loop.}
\label{fig:zee}
\end{figure}
Without loss of generality, we work in the Higgs basis where only the scalar doublet $H_1$ takes a VEV $v = 246~\text{GeV}$. The Yukawa Lagrangian of the model reads
\begin{equation}
    -\mathcal{L}_{\rm Zee}=\overline{L}\left(y^\dagger_e H_1+\Gamma^\dagger_e H_2\right)e_R+ \overline{\tilde L}f L h^++\text{ H.c.}\,,
\end{equation} 
where $e_R$ are the right-handed charged lepton singlets. Here $y_e$ can be taken to be diagonal, i.e., we are in the charged lepton mass basis, with $M_e=y^\dagger_e v/\sqrt{2}$, whereas $\Gamma_e$ is a general complex $n_L \times n_L$ matrix and $f$ is a complex $n_L \times n_L$ anti-symmetric matrix, $f^T=-f$. The scalar potential of the Zee model includes the LNV trilinear term:
\begin{equation}
    V\subset \mu \tilde H_1^\dagger H_2 h^- +\text{ H.c.}\,,
\end{equation}
which, without loss of
generality, can be chosen to be real and positive by a field redefinition of the singlet $h^+$. After the EW SSB, it induces a mixing between the singly-charged scalars of the doublet $H_2$ and the singlet $h^+$, 
\begin{equation}
    \sin2\varphi=\dfrac{\sqrt{2}v\mu}{m_{h^+_2}^2-m_{h^+_1}^2}\,,
\end{equation}
where $h_1^+$ and $h_2^+$ are the mass eigenstates, with $m_{h^+_2}>m_{h^+_1}$. The Majorana neutrino mass matrix reads: 
\begin{equation}
    m_\nu=\dfrac{\sin2\varphi}{16\pi^2}\left( f M_e\Gamma_e +\Gamma^T_e M_e^T f^T\right)\log\left( \dfrac{m_{h^+_2}^2}{m_{h^+_1}^2}\right)\,.
\end{equation}
This mass matrix can be rewritten in the form of Eq.~\eqref{eq:mnuSS} with the identifications
\begin{equation}
    \mathcal{Y}=\begin{pmatrix}
        \hat f\\
        \Gamma_e
    \end{pmatrix}\,,\qquad {\mathcal{M}}=\begin{pmatrix}
        0&M_e\\
        M_e^T&0
    \end{pmatrix} \,,
\end{equation}
where ${M_e}$ is the diagonal matrix of charged lepton masses and we have absorbed the global factors by defining
\begin{equation}
    \hat f=\dfrac{\sin2\varphi}{16\pi^2}\,\log\left( \dfrac{m_{h^+_2}^2}{m_{h^+_1}^2}\right)\, f^T\,.
\end{equation}
The explicit form of the matrix $V$ in $D_{\mathcal{M}}=V^\ast \mathcal{M} V^\dagger$ is given in Eq.~\eqref{eq:Vgen}. This allows for the use of the GCI parametrisation, Eq.~\eqref{eq:CIgen}. 
However, in this scenario, the matrix $\mathcal{R}$ has less degrees of freedom than an arbitrary semi-orthogonal complex matrix, because it must realize the antisymmetric structure of the Yukawa matrix associated with the singly-charged scalar $h^+$, i.e., $f^T=-f$.

In the following, we show that it is possible to find an explicit parametrisation of the $\mathcal{R}$ matrix that guarantees the antisymmetric condition of the Yukawa matrix $f$. For this, it is convenient to write $\mathcal{R}$ as
\begin{equation}
    \mathcal{R}=\begin{pmatrix}
        R_1\\
        R_2
    \end{pmatrix},
\end{equation}    
where $R_1$ and $R_2$ are $n_L \times n_L$ complex matrices, with $n_L=3$. In this way, the orthogonality condition $\mathcal{R}^T \mathcal{R} = \mathbb{I}_{3}$ reads 
\begin{equation}
\label{eq:zee_orthogonal_cond}
    R_1^T R_1 + R_2^T R_2 = \mathbb{I}_{3}\,.
\end{equation}
In addition, from the GCI parametrisation, considering the upper block part, we get
\begin{equation}\label{eq:CIZee}
\hat f = \frac{-i}{\sqrt{2}} M_e^{-1/2}\, (R_1 + i R_2) \,D_{\sqrt{m}}\,U^\dagger\,.
\end{equation}
Therefore, 
\begin{equation}
\label{eq:zee_R1_R2_relation}
    R_1 + i R_2 = A := i \sqrt{2} \, M_e^{1/2}\, \hat f \, U \, D^{-1/2}_{m}\,,
\end{equation}
where we have introduced the matrix $A$ and assumed that $D_{m}$ is invertible. 
Transposing this equation, it is immediate to see that
\begin{equation}
R_1^T R_1 = A^T A - i (R_2^T A + A^T R_2) - R_2^T R_2\,.
\end{equation}
Notably, from this equation and the orthogonality condition in Eq.~\eqref{eq:zee_orthogonal_cond}, we arrive at a condition which is linear in $R_2$, namely,
\begin{equation}
\label{eq:zee_R2_equation}
    R_2^T \, A + A^T \, R_2 = i \left(\mathbb{I}_{3} - A^T A \right)\,.
\end{equation}

Then it is possible to choose 6 free (complex) parameters and obtain analytical expressions for the remaining ones in the following way. Two of the three Yukawa couplings in $\hat f$ can be freely chosen, while the third one is fixed by the condition that $\hat f$, being of rank 2, has a null eigenvector~\cite{Felkl:2021qdn}. Specifically, writing $\hat f$ as
\begin{equation}
    \hat f=\begin{pmatrix}
        0& \hat f_3 & - \hat f_2\\
        -\hat f_3 &0& \hat f_1\\
        \hat f_2 & -\hat f_1 & 0
    \end{pmatrix}\, ,
\end{equation}
$\hat v=(\hat f_1, \hat f_2, \hat f_3)^T$ is a null eigenvector of $\hat f$, which satisfies 
\begin{equation}
\label{eq:zee_v_condition}
\hat v^T \, m_\nu \, \hat v = 0\,, 
\end{equation}
leading to a quadratic equation that fixes one of the elements of $\hat f$ in terms of the other two~\cite{Felkl:2021qdn}. Moreover, once $\hat f$ is fixed, the matrix $A$ is completely determined (for a given light neutrino mass matrix) and therefore Eq.~\eqref{eq:zee_R2_equation} leads to a linear system of 6 equations for the 9 elements of $R_2$. Actually, given that $\hat f$ is not invertible, neither is $A$, and one can show that there are only five independent linear equations for the elements of $R_2$. The linear system can always be solved analytically, yielding 5 elements of $R_2$ in terms of the other four that are chosen as free parameters. In turn, $R_1$ is determined from Eq.~\eqref{eq:zee_R1_R2_relation} (which can also be used to choose a different set of free parameters). 

In order to write explicitly an independent set of linear equations for the elements of $R_2$, note that from $\hat f \, \hat v = 0$ it follows that 
\begin{equation}
    A (D_{\sqrt m} \, U^\dagger \hat v)=0.
\end{equation}
From this condition it is possible to write the first column of $A$ as a linear combination of the second and third columns, namely 
\begin{equation}
    A_{i1}= \lambda_2 A_{i2} + \lambda_3 A_{i3},
\end{equation}
with $\lambda_{2,3}$ determined from $\hat f$ and $m_\nu$:
\begin{align}
    \lambda_2 & = - \frac{\sum_i \sqrt{m_2} \, U_{i2}^* \, \hat f_i}{\sum_i \sqrt{m_1} \, U_{i1}^* \, \hat f_i} \, ,\nonumber\\
    \lambda_3 & = - \frac{\sum_i \sqrt{m_3}\, U_{i3}^* \, \hat f_i}{\sum_i \sqrt{m_1} \, U_{i1}^* \, \hat f_i}  \,,
\end{align}
for $\sum_i \sqrt{m_1} \, U_{i1}^* \, \hat f_i\neq 0$.
Then, a proper system of five linear independent equations for the elements of $R_2 := [r_{ij}]$, writing also $A:=[a_{ij}]$ and $i \left(\mathbb{I}_{3} - A^T A \right):=[b_{ij}]=[b_{ji}]$, is 
\begin{equation}
\left\{ \quad
\begin{aligned}
 2 \sum_j \left(\lambda_2 \, a_{j2} + \lambda_3 \, a_{j3}\right) \, r_{j1} &= b_{11} \\
  \sum_j a_{j3} \, r_{j1} + \left( \lambda_2  \, a_{j2} + \lambda_3 \, a_{j3} \right) \, r_{j3} &= b_{13} \\ 
2 \sum_j  a_{j2}\, r_{j2} &= b_{22} \\
\sum_j a_{j3} \,r_{j2} + a_{j2}\, r_{j3} &= b_{23} \\
2 \sum_j  a_{j3} \,r_{j3} &= b_{33} \, ,
\label{eq:zee_r2_elements_eq}
\end{aligned}
\right.
\end{equation}
where, if not specified, the summations go from $i (j)=1$ to $i (j)=3$.\footnote{The equation $\sum_j a_{j2}\,r_{j1}+(\lambda_2 a_{j2}+\lambda_3 a_{j3})\,r_{j2} = b_{12}$ is a linear combination of the other ones and therefore it has been omitted.}
Here it is possible to choose different sets of four free parameters and express the remaining ones in terms of them, typically by solving smaller subsystems at a time. For example, choosing $r_{11}, r_{21}, r_{12}$ and $r_{13}$ as free parameters, the solution is: 

\begin{align}
r_{31} & = \frac{1}{\lambda_2 \, a_{32} + \lambda_3 \, a_{33}} \left( \frac{b_{11}}{2} - \sum_{j=1}^2 \left(\lambda_2 \, a_{j2} + \lambda_3 \, a_{j3}\right) \, r_{j1} \right) \, \notag, \\
    \vspace{0.1cm}\\
    \begin{pmatrix}
        r_{23} \\
        r_{33}
    \end{pmatrix}
    & = \frac{1}{\lambda_{2} \, d}
    \begin{pmatrix} 
        a_{33}  \quad &  -(\lambda_2 a_{32} + \lambda_3 a_{33})\\[4pt]
        -a_{23} \quad & \lambda_2 a_{22} + \lambda_3 a_{23}
    \end{pmatrix}\,\times\nonumber\\
    &\begin{pmatrix}
        b_{13} - \sum_{j=1}^3 a_{j3} \, r_{j1} - \left( \lambda_2  \, a_{12}  + \lambda_3 \, a_{13}\right) \, r_{13} \\[4pt]
        b_{33}/2 - a_{13} \, r_{13}
    \end{pmatrix} \,, \notag \\
    \vspace{0.1cm}\\
    \begin{pmatrix}
        r_{22} \\[4pt]
        r_{32}
    \end{pmatrix}
    & = \frac{1}{d}
    \begin{pmatrix} 
        a_{33} &  -a_{32} \\
        -a_{23} & a_{22}
    \end{pmatrix} \,
    \begin{pmatrix}
    b_{22}/2 - a_{12} \, r_{12} \\
        b_{23} - a_{13} \, r_{12} - \sum_{j=1}^3 a_{j2} \, r_{j3}  
    \end{pmatrix} \,, 
\end{align}
with $d = a_{22} \, a_{33} - a_{23} \, a_{32}$. Finally, $R_1$ can be obtained from $R_2$ using Eq.~\eqref{eq:zee_R1_R2_relation} and we arrive at an explicit parametrisation for both $\hat f$ (see Eq.~\eqref{eq:CIZee}) and $\Gamma_e$,
\begin{equation}\label{eq:CIZeeG}
\Gamma_e = \frac{1}{\sqrt{2}} M_e^{-1/2}\, (i R_1 + R_2) \,D_{\sqrt{m}}\,U^\dagger\,,
\end{equation}
which results from considering the lower block part of the GCI parametrisation. 

This procedure gives a parametrisation of the Zee model and other models with an antisymmetric Yukawa matrix, with analytical expressions for all the parameters in terms of a chosen set of free parameters. Note that once the Yukawa couplings in $f$ are fixed, instead of following the GCI procedure outlined above, it would also be possible to set and solve analytically a linear system of equations directly for the Yukawa couplings in $\Gamma_e$. Also notice that a different parametrisation was proposed in Ref.~\cite{Machado:2017flo}\footnote{Note that in the Zee-Wolfenstein model \cite{Wolfenstein:1980sy} only one scalar doublet
couples to the SM leptons (imposed via a discrete symmetry), and therefore $\rm{Tr}\, (m_\nu) = 0$.  However, this possibility is now excluded by data. Therefore, in general, both scalar doublets couple to the leptons and $\rm{Tr}\, (m_\nu) \neq 0$.} and it was applied to phenomenological studies in Refs.~\cite{Heeck:2023iqc, Dev:2023nha}. Further insight into the relationship among the different parameters of the Zee model and observational constraints, particularly on charged lepton flavour violation, was given in Ref.~\cite{Chen:2025thp}. Since different parametrisations may have different virtues, comparing and applying these approaches could be an interesting direction for future work.

\section{The complete mass matrix with right-handed neutrinos: Discussion and Outlook} \label{sec:CIgen}

It is suggestive to rewrite the complete neutrino mass matrix including right-handed neutrinos, \(M_\nu\) in Eq.~\eqref{eq:MMSS2}, in the form
\begin{equation}\label{ECIMassMatrix2}
M_\nu=\widetilde{\mathcal Y}^T\,\widetilde{\mathcal M}\,\widetilde{\mathcal Y},
\end{equation}
with
\begin{equation}  \label{eq:Ytilde}
\widetilde{\mathcal Y}=
\begin{pmatrix} 
Y^T & 0_{\,n_R\times n_R}\\[6pt]
0_{\,n_R\times n_L} & \mathbb I_{n_R}
\end{pmatrix}\in\mathbb C^{2n_R\times(n_L+n_R)}
\end{equation}
and
\begin{equation} \label{eq:tildeM}
\widetilde{\mathcal M}= 
\begin{pmatrix} 
0_{n_R} & v/{\sqrt{2}}\,\mathbb I_{n_R}\\[6pt]
v/{\sqrt{2}}\,\mathbb I_{n_R} & m_R
\end{pmatrix}\in\mathbb C^{2n_R\times 2n_R}\,,
\end{equation}
where we have written quantities with a tilde to point out that we are working with the complete mass matrix. Note that \(\widetilde{\mathcal Y}\) is rectangular (unless \(n_L=n_R\)) and \(\widetilde{\mathcal M}\)
is symmetric. In principle, it is possible to formally write a GCI structure in this case, similar to Eq.~\eqref{eq:CIgen}, 
\begin{equation}  \label{eq:tildeYa}
\mathcal{\widetilde Y}=V^\dagger D^{-1/2}_{\mathcal{\widetilde M}}\,\mathcal{\widetilde R}\,D^{1/2}_{M}\,W^\dagger\,,
\end{equation}
where we used a Autonne-Takagi decomposition for the $(n_L+n_R)\times(n_L+n_R)$ complete neutrino mass matrix, 
\begin{equation}
    D_{M}=\text{diag}(m_1,\,\ldots,\,m_{n_L+n_R})=W^T M_\nu W\,.
\end{equation}
Here $V$ and $W$ are $2n_R\times 2n_R$ and  $(n_L+n_R)\times(n_L+n_R)$ unitary matrices, respectively, and $\mathcal{\widetilde R}$ is a $2 n_R \times (n_L+n_R)$ complex orthogonal matrix, $\mathcal{\widetilde R}^T \mathcal{\widetilde R}=\mathbb I_{n_L+n_R}$. Note that $\mathcal{\widetilde Y}$ is not general, as it has many zeroes and the identity matrix, therefore $\mathcal{\widetilde R}$ is not arbitrary either. A natural question emerges:\\
\vspace{0.1cm}
\[
\boxed{
\begin{minipage}{0.85\linewidth}
\emph{Can one use $\widetilde{\mathcal{Y}}$ to obtain a parametrisation for $Y$ valid outside of the seesaw limit, or one that uses some heavy-light mixings as input?}
\end{minipage}
}
\]
\noindent
\vspace{0.1cm}\\
Although deriving a generic parametrisation valid outside of the seesaw limit using this approach would definitely be very interesting, it is beyond the scope of this work. For other parametrisations in models with right-handed neutrinos, which are exact and valid outside of the seesaw limit, see Refs.~\cite{Blennow:2011vn,Donini:2012tt}. In the next section and in App.~\ref{app:oldCI} we discuss the seesaw limit, which of course should give rise to the standard CI at leading order.

\subsection{The standard Casas-Ibarra parametrisation in the seesaw limit}

In the seesaw limit, $m_D \ll m_R$ (small light--heavy mixing), the complete neutral fermion mass matrix $M_\nu$ can be approximately block-diagonalised to $D_M=\mathrm{diag}(D_\nu,D_R)$ by the full unitary matrix $W$. It is convenient to factor the later as
\begin{equation}
W=\Omega\,(U\oplus \mathbb I_{n_R})\,,
\end{equation}
where $U$ is the PMNS matrix and $\Omega$ encodes the light--heavy mixing. The corresponding complete unitary mixing matrix that diagonalises $M_\nu$ is
\begin{align}
&W \;\simeq\;
\begin{pmatrix}
(\mathbb I_{n_L}-\tfrac{1}{2}\Theta\Theta^{\dagger})\,U & \Theta \\
-\Theta^{\dagger}U & \mathbb I_{n_R}-\tfrac{1}{2}\,\Theta^{\dagger}\Theta\,
\end{pmatrix}
\;+\;\mathcal O(\Theta^{3})\,,\nonumber\\
&\Theta = m_D\,m_R^{-1}\in \mathbb{C}^{n_L \times n_R},
\end{align}
This gives rise to mass matrices in the diagonal of the form:
\begin{align}
&m_\text{light} \simeq \, m_D m_R^{-1} m_D^T \in \mathbb{C}^{n_L\times n_L}\,,\nonumber\\
&m_\text{heavy} \simeq m_R \in \mathbb{C}^{n_R\times n_R}\,.
\end{align}
In App.~\ref{app:oldCI} we explicitly show how the standard CI parametrisation,
\begin{equation} \label{eq:oldCI}
\boxed{~
m_D \;=\; U^*\,D_\nu^{1/2}\, R_{\ell}\, D_R^{1/2} \,,
~}
\end{equation}
valid up to $\mathcal{O}(\Theta^2)$, may be derived in the seesaw limit, as expected, from Eq.~\eqref{eq:tildeYa}. Here $R_\ell$ encodes the remaining freedom:
\begin{equation}
R_{\ell} \in \mathbb C^{\,n_L\times n_R},\qquad
R_{\ell}\,R_{\ell}^{T}=\mathbb I_{n_L}\quad\text{if }~n_R\ge n_L,
\end{equation}
while for $n_R<n_L$ one has $R_{\ell}^{T}R_{\ell}=\mathbb I_{n_R}$.

\section{Conclusions} \label{sec:conc}

We have shown that a simple generalisation of the Casas–Ibarra (GCI) parametrisation can be applied to all Majorana neutrino mass models. In particular, it can be straightforwardly applied to scenarios where there are several contributions to the Majorana neutrino mass matrix, with both diagonal and off-diagonal terms in Yukawa space. This is highly relevant to perform phenomenological studies, and, in particular, it allows to control certain somewhat fine-tuned regions of the parameter space when doing a numerical scan, like those with large Yukawa hierarchies, which are governed by the imaginary parts of the complex angles of the (semi-)orthogonal $\mathcal{R}$ matrix.

The approach, in turn, provides a unified framework for classifying neutrino mass models according to the structure of the neutrino mass matrix. This strategy has highlighted the lack of certain types of models, naturally leading to the proposal of an extended version of the Scotogenic Model whose phenomenology may be explored elsewhere. In particular, we have discussed different tree-level (one loop-level) examples, including the seesaw (the Scotogenic Model), the linear seesaw (the Generalised Scotogenic Model), and the linear plus inverse seesaw (the newly proposed Extended Scotogenic Model). 

We have also provided ready-to-use expressions for the Yukawa matrices in several well-known models. In particular, for the Zee model we have derived the constraints on the $\mathcal{R}$ matrix and provided an explicit parametrisation. With our procedure, using Eq.~\eqref{eq:tildeYa}, it may also be possible to apply the parametrisation outside of the seesaw limit, which is left for future work.

\vspace{0.5cm} \noindent
{\bf \underline{Note Added} \vspace{0.2cm}\\}
While this work was being finalised, Ref.~\cite{Herrero-Brocal:2025zpb} appeared on the \emph{arXiv}, proposing a generalisation of the Casas--Ibarra parametrisation and demonstrating its equivalence to the master parametrisation of Refs.~\cite{Cordero-Carrion:2018xre,Cordero-Carrion:2019qtu}. We emphasise that Ref.~\cite{Herrero-Brocal:2025zpb} pursues a distinct aim: it focuses on conditions that reduce the number of free parameters, whereas our emphasis is on generality and on applications to widely studied models. The new model we propose is, in fact, motivated by this broader perspective.

\begin{acknowledgments}
We are grateful for useful discussions with Pilar Hernández, Jacobo López-Pavón, Nuria Rius and Alejandro Ibarra. We also acknowledge discussing with Avelino Vicente at FLASY 2025 regarding our insight of a generalisation of the Casas–Ibarra parametrisation. SM would like to thank Federico Mescia for the kind hospitality at the \emph{Laboratori Nazionali di Frascati} (LNF-INFN) during the early stages of this work. This work is partially supported by the Spanish \emph{Agencia Estatal de Investigación} MICINN/AEI (10.13039/501100011033) grant PID2023-148162NB-C21 and the \emph{Severo Ochoa} project MCIU/AEI CEX2023-001292-S. JHG is supported by the \emph{Consolidación Investigadora Grant CNS2022-135592}, funded also by \emph{European Union NextGenerationEU / PRTR}, and the Generalitat Valenciana Plan GenT Excellence Program (CIESGT2024/7). SM acknowledges financial support from the \emph{Consolidación Investigadora Grant CNS2022-136005}. JR thanks the support by a grant Consolidar-2023-2027 from the Secretar\'ia de Ciencia y Tecnolog\'ia (SeCyT) de la Universidad Nacional de C\'ordoba (UNC) and also support from the Consejo Nacional de Investigaciones Cient\'ificas y T\'ecnicas \mbox{(CONICET)}, Argentina. DV is supported by a McDonald Institute Theory Fellowship funded from the Canada First Research Excellence Fund through the Arthur B. McDonald Canadian Astroparticle Physics Research Institute, and a Subatomic Physics Discovery Grant (individual) from the Natural Sciences and Engineering Research Council of Canada.  All Feynman diagrams were generated using the Ti\textit{k}Z-Feynman package for \LaTeX~\cite{Ellis:2016jkw}.

\end{acknowledgments}

\appendix

\section{Derivation of the standard Casas-Ibarra from the general one} \label{app:oldCI}

To make contact with the usual CI parametrisation, it is instructive to analyse the seesaw limit of the general relation given in Eq.~\eqref{eq:tildeYa}, 
\begin{equation}  \label{eq:tildeYb}
\mathcal{\widetilde Y}=V^\dagger D^{-1/2}_{\mathcal{\widetilde M}}\,\mathcal{\widetilde R}\,D^{1/2}_{M}\,W^\dagger\,,
\end{equation}
where $D_M=\mathrm{diag}(D_\nu,D_R)$. It is convenient to factor the full diagonalisation matrix as
\begin{equation}
W=\Omega\,(U\oplus \mathbb I_{n_R})\,,
\end{equation}
where $U$ is the PMNS matrix and $\Omega$ encodes the light--heavy mixing, so that: 
\begin{equation}\label{eq:CIgen-explicit}
\boxed{
\widetilde{\mathcal Y}
= X\,\widetilde{\mathcal R}\,D_M^{1/2}\,(U^\dagger\oplus\mathbb I_{n_R})\,\Omega^\dagger,
\qquad
X\equiv V^\dagger D_{\widetilde{\mathcal M}}^{-1/2}.
}
\end{equation}
\onecolumngrid
We partition the matrices as
\begin{equation}
X=\begin{pmatrix}X_{aa}&X_{ab}\\[2pt]X_{ba}&X_{bb}\end{pmatrix},\quad
\widetilde{\mathcal R}=\begin{pmatrix}A&B\\[2pt]C&D\end{pmatrix},\quad
\Omega=\begin{pmatrix}K&R\\[2pt]S&T\end{pmatrix}\,.
\end{equation}
Then
\begin{equation}
\widetilde{\mathcal Y}
= \begin{pmatrix} A_1 & B_1\\[2pt] C_1 & D_1 \end{pmatrix}
\begin{pmatrix}
K^\dagger & S^\dagger\\[2pt]
R^\dagger & T^\dagger
\end{pmatrix}
=
\begin{pmatrix}
A_1 K^\dagger + B_1 R^\dagger &\; A_1 S^\dagger + B_1 T^\dagger\\[4pt]
C_1 K^\dagger + D_1 R^\dagger &\; C_1 S^\dagger + D_1 T^\dagger
\end{pmatrix}\,,
\end{equation}
where we defined the intermediate blocks
\begin{align} 
A_1 &\equiv (X_{aa}A+X_{ab}C)\,D_\nu^{1/2}\,U^\dagger, &
B_1 &\equiv (X_{aa}B+X_{ab}D)\,D_R^{1/2},\label{eq:inter}\\
C_1 &\equiv (X_{ba}A+X_{bb}C)\,D_\nu^{1/2}\,U^\dagger, \label{eq:intera}& 
D_1 &\equiv (X_{ba}B+X_{bb}D)\,D_R^{1/2}.
\end{align}
Equating with \(\widetilde{\mathcal Y}=\mathrm{diag}(Y^T,\mathbb I_{n_R})\) yields
\begin{align}
&Y^T \;=\; A_1 K^\dagger + B_1 R^\dagger, \label{eq:TL-gen}\\
&0 \;=\; A_1 S^\dagger + B_1 T^\dagger, \label{eq:TR-gen}\\
&0 \;=\; C_1 K^\dagger + D_1 R^\dagger, \label{eq:BL-gen}\\
&\mathbb I \;=\; C_1 S^\dagger + D_1 T^\dagger. \label{eq:BR-gen}
\end{align}
In the seesaw regime \(\|m_D\|\ll\|m_R\|\), we have
\begin{equation}
\Omega \simeq
\begin{pmatrix}
\mathbb I_{n_L}-\tfrac12\Theta\Theta^\dagger & \Theta\\[2pt]
-\Theta^\dagger & \mathbb I_{n_R}-\tfrac12\Theta^\dagger\Theta
\end{pmatrix},
\qquad
\Theta\equiv m_D\,D_R^{-1}\in\mathbb C^{n_L\times n_R},
\end{equation}
so that, to leading order,
\begin{equation}
K^\dagger=\mathbb I+\mathcal O(\Theta^2),\quad
T^\dagger=\mathbb I+\mathcal O(\Theta^2),\quad
R^\dagger=\Theta^\dagger+\mathcal O(\Theta^3),\quad
S^\dagger=-\Theta+\mathcal O(\Theta^3).
\end{equation}
Equations \eqref{eq:BL-gen} -- \eqref{eq:BR-gen} imply
\begin{equation}
C_1=\mathcal O(\Theta),\qquad D_1=\mathbb I+\mathcal O(\Theta^2),
\end{equation}
and $B_1=0$ at zeroth order from Eq.~\eqref{eq:TR-gen}.
Solving the $B_1$ and $D_1$ equations, we obtain at \(\Theta^0\)
\begin{equation}
C \;=\; -\,X_{bb}^{-1}X_{ba}\,A \;+\;\mathcal O(\Theta),
\qquad
D\,D_R^{1/2} \;=\; S_b^{-1} \;+\;\mathcal O(\Theta^2),
\end{equation}
where the Schur complements of \(X\) are (we assume that $X_{aa}, X_{bb}$ are invertible)
\begin{equation}
S_a \equiv X_{aa}-X_{ab}X_{bb}^{-1}X_{ba},
\qquad
S_b \equiv X_{bb}-X_{ba}X_{aa}^{-1}X_{ab}.
\end{equation}
With these, we obtain
\begin{equation}
A_1 = S_a\,A\,D_\nu^{1/2}\,U^\dagger + \mathcal O(\Theta),\qquad
B_1 = X_{aa}B\,D_R^{1/2} + X_{ab}S_b^{-1} + \mathcal O(\Theta).
\end{equation}
Using Eq.~\eqref{eq:TL-gen}, we get
\begin{equation}
~Y^T \;=\; S_a\,A\,D_\nu^{1/2}\,U^\dagger
\;+\; \big(X_{aa}B\,D_R^{1/2}+X_{ab}S_b^{-1}\big)\,\Theta^\dagger \,,
\label{eq:Y1-leading}
\end{equation}
where the second term is ${\mathcal O(\Theta^2)}$. From the Autonne-Takagi decomposition of \(\widetilde{\mathcal M}\) given in Eq.~\eqref{eq:tildeM},
\begin{equation}
D_{\widetilde{\mathcal M}}
\simeq \mathrm{diag}\!\Big(\tfrac{v^2}{2}\,D_R^{-1},\,D_R\Big),\qquad
V\simeq
\begin{pmatrix}\mathbb I_{n_R} -\tfrac12\Phi\Phi^\dagger& \Phi\\[2pt]-\Phi^T & \mathbb I_{n_R}-\tfrac12\Phi^T\Phi^*\end{pmatrix},
\qquad
\Phi\equiv \tfrac{v}{\sqrt2}\,D_R^{-1},
\end{equation}
so neglecting $\mathcal{O}(\Phi^2)$ terms,
\begin{equation}
X \equiv V^\dagger D_{\widetilde{\mathcal M}}^{-1/2}
\simeq
\begin{pmatrix}
\alpha\,\mathbb I & -\,\beta\,\Phi\\[2pt]
\alpha\,\Phi & \beta\,\mathbb I
\end{pmatrix},
\qquad
\alpha\equiv \frac{\sqrt2\,D_R^{1/2}}{v},\quad
\beta\equiv D_R^{-1/2}.
\end{equation}
Then, as $X_{aa}, X_{bb}$ are invertible because $\alpha, \beta$ are ($D_R>0$):
\begin{equation}
S_a
= X_{aa}-X_{ab}X_{bb}^{-1}X_{ba}
= \alpha\big(\mathbb I+\Phi^2\big)
= \frac{\sqrt2\,D_R^{1/2}}{v}\,\Big[\mathbb I+\mathcal O\!\Big(\frac{v^2}{2}\,D_R^{-2}\Big)\Big].
\end{equation}
Dropping the \(\mathcal O(\Theta^2)\) correction in Eq.~\eqref{eq:Y1-leading} and choosing
\begin{equation}
~A\equiv R_\ell^{\,T},\qquad R_\ell\ \text{complex orthogonal}\,,
\end{equation}
with $R_\ell R_\ell^T=\mathbb I_{n_L}$ for $n_R\ge n_L$ and
$R_\ell^T R_\ell=\mathbb I_{n_R}$ for $n_R<n_L$,
we obtain:
\begin{equation}
Y^T U \;\simeq\; \frac{\sqrt{2}}{v}\,D_R^{1/2}\,R_\ell^{\,T}\,D_\nu^{1/2}
\qquad\Longrightarrow\qquad
\boxed{~m_D \;\simeq\;\,U^*\,D_\nu^{1/2}\,R_\ell\,D_R^{1/2}~,}
\end{equation}
which is the familiar CI parametrisation, valid up to 
\(\mathcal O(\Theta^2)\).
\twocolumngrid

\bibliographystyle{utphys}
\bibliography{refs}

\end{document}